\documentclass[11pt]{article}

\usepackage[preprint]{acl}

\usepackage{times}
\usepackage{latexsym}

\usepackage[T1]{fontenc}

\usepackage[utf8]{inputenc}

\usepackage{microtype}

\usepackage{inconsolata}

\usepackage{graphicx}
\usepackage{enumitem}

\usepackage{amsmath}
\usepackage{amssymb}
\usepackage{booktabs}
\usepackage{multirow}
\usepackage{float}
\usepackage[table]{xcolor}

\title{SAILRec: Steering LLM Attention to Dual-Side Semantically Aligned Collaborative Embeddings for Recommendation}

\author{
  \textbf{Xi Wu},
  \textbf{Jiale Wang},
  \textbf{Zihan Wang},
  \textbf{Yichen Gao},
\\
  \textbf{Xiaocui Yang},
  \textbf{Shi Feng},
  \textbf{Daling Wang},
  \textbf{Yifei Zhang}
\\
\\
  Northeastern University, China
\\
  {\small
  \begin{tabular}{c}
  \texttt{wux2@mails.neu.edu.cn, wangjiale12356@mail.ustc.edu.cn, wzh1998921@gmail.com, } \\
  \texttt{gaoyc3@mails.neu.edu.cn,\{yangxiaocui, fengshi, wangdaling, zhangyifei1\}@cse.neu.edu.cn}
  \end{tabular}
  }
}

\begin{document}
\maketitle

\begin{abstract}
Recent LLM-based recommenders enhance language models with collaborative embeddings from user-item interactions, but making such embeddings available does not ensure their proper use during inference. Through a diagnostic attention analysis, we find that the utilization of collaborative embeddings is depth-dependent and alignment-sensitive, suggesting that LLMs need to balance their internal semantic knowledge with external collaborative knowledge. To address this issue, we propose SAILRec, an LLM-based recommender that improves this balance through dual-side semantic alignment and hierarchical attention steering. The former aligns item-side embeddings with item-text semantics and user-side embeddings with codebook-based semantic profiles, while the latter suppresses premature shallow-layer collaborative interference and strengthens collaborative evidence in deeper decision layers. Experiments on MovieLens-1M and Amazon-Book show that SAILRec consistently outperforms representative baselines, with ablation and masking analyses validating its key designs.
\end{abstract}

\section{Introduction}

Large language models (LLMs) bring strong semantic understanding and instruction-following abilities to recommendation~\citep{survey-llmrec-2023-1,survey-llmrec-2024-1,survey-llmrec-2024-2}, but they do not naturally model collaborative knowledge hidden in interaction histories. Recent methods therefore inject collaborative embeddings derived from collaborative filtering models (Collab.) into LLMs as soft tokens, as in CoLLM~\citep{collm}, but the collaborative knowledge encoded in these embeddings remains separated from the semantic space of LLMs. SeLLa-Rec~\citep{sellarec} shows that semantic alignment can mitigate this gap through item-side distillation and contrastive alignment, indicating that collaborative knowledge should be made semantically compatible with the LLM before being injected. Nevertheless, existing methods mainly focus on constructing and injecting collaborative embeddings, while paying less attention to how LLMs actually balance and use these collaborative signals with their internal semantic knowledge during inference.

\begin{figure*}[t]
    \centering
    \includegraphics[width=1\textwidth]{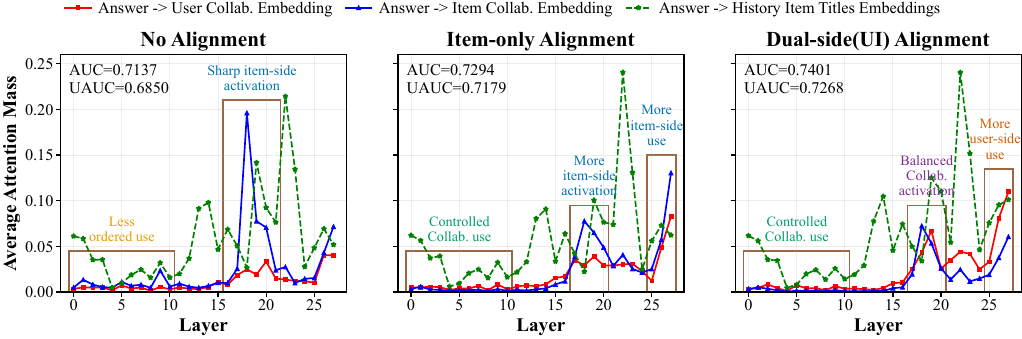}
    \caption{Mean attention from the answer position to key token groups under different semantic alignment settings on MovieLens-1M. Collaborative embedding utilization is depth-dependent and alignment-sensitive. No alignment shows less ordered early use and sharp item-side activation, item-side-only alignment improves late-layer accessibility of collaborative signals, and dual-side alignment achieves better performance with more balanced collaborative use and increased top-layer user-side attention.}
    \label{fig:mean_attention}
\end{figure*}

To examine how LLMs use collaborative embeddings during inference, we conduct a diagnostic experiment on MovieLens-1M. Since scaled dot-product attention controls value aggregation from different token groups and is widely used to probe Transformer behavior~\citep{attention,bert-attention,quantify-attention}, we use attention scores as diagnostic signals. Prior layer-wise analyses suggest that shallow Transformer layers mainly encode lexical, local syntactic, and contextual features, while deeper layers integrate higher-level semantic and task-relevant information~\citep{bert-pipeline}. Based on this view, we adopt a CoLLM-style injection structure and measure answer-position attention to collaborative embeddings on the user-side, collaborative embeddings on the item-side, and historical embeddings of the item-title, as shown in Figure~\ref{fig:mean_attention}. We do not assume that a specific attention curve is inherently optimal. Instead, we use it to diagnose how collaborative embeddings are internally utilized. The results reveal two observations. First, collaborative embedding utilization is depth-dependent. Their attention remains low in shallow layers and becomes more active in middle-to-deep layers, suggesting that collaborative knowledge is more likely to contribute to preference matching and decision-stage prediction than to early textual encoding. Second, semantic alignment significantly changes the utilization pattern of collaborative embeddings and is associated with better recommendation performance. These observations suggest that effective use of collaborative embeddings requires addressing two questions: whether they are understandable to the LLM semantic space, and when they should affect Transformer inference.

Based on these findings, we propose SAILRec, short for \textbf{S}teering \textbf{A}ttention to collaborative embeddings \textbf{I}n \textbf{L}LM \textbf{Rec}ommendation. It is an LLM-based recommender that improves collaborative embedding utilization. SAILRec addresses the above two questions from complementary aspects. First, it performs dual-side semantic alignment to improve the semantic accessibility of user-side and item-side collaborative embeddings. The item side is aligned with LLM semantic representations of item texts, while the user side is aligned with codebook-based semantic profiles from historical interactions. Second, SAILRec introduces hierarchical attention steering to control when collaborative knowledge is used across Transformer layers. This mechanism follows the tendency that collaborative embeddings become more active in middle-to-deep layers. It suppresses premature collaborative interference in shallow layers, preserves natural interaction in middle layers, and strengthens collaborative evidence in deep layers for preference matching. By combining both designs, SAILRec makes external collaborative knowledge more accessible to the LLM semantic space and more appropriately involved in recommendation decisions. Our contributions can be summarized as follows:
\begin{itemize}[
    leftmargin=*,
    itemsep=1pt,
    parsep=0pt,
    topsep=1pt,
]
\item We provide a diagnostic analysis showing that injected collaborative embeddings are utilized in a depth-dependent and alignment-sensitive manner, motivating the need to improve both their semantic accessibility and layer-wise usage.

\item We propose SAILRec, which combines dual-side semantic alignment and hierarchical attention steering to improve the semantic accessibility and layer-wise utilization of Collab. embeddings.

\item We conduct experiments on two public datasets, showing that SAILRec outperforms representative baselines and that ablation and masking analyses validate its key designs.
\end{itemize}

\section{Related Work}
\subsection{LLM as Recommender}
LLM-only recommendation verbalizes user histories, candidate items, and task objectives, enabling LLMs to perform preference prediction, ranking, or next-item prediction through prompting, instruction tuning, or lightweight adaptation~\citep{p5,chatrec}. TALLRec~\citep{tallrec} aligns LLMs with recommendation tasks through Alpaca tuning and rec-tuning, while RecRanker~\citep{rec-ranker} trains LLMs as top-$K$ rankers with user sampling, position-shifted prompts, and hybrid pointwise, pairwise, and listwise ranking. RecLM~\citep{rec-lm} introduces recommendation instruction tuning with reinforcement learning rewards to capture diverse user preferences. Later studies move from task tuning to structural adaptation. CTR-Sink~\citep{ctr-sink} and HatLLM~\citep{hatllm} examine attention mechanisms, addressing semantic fragmentation in behavior sequences and excessive within-item attention. Overall, these studies demonstrate the potential of LLMs as semantic recommenders, but remain limited in modeling collaborative signals encoded in user/item IDs and interaction structures.

\subsection{LLM with Collaborative Knowledge}
To mitigate the limited ability of LLM-only recommenders to capture collaborative signals, recent studies integrate knowledge from collaborative filtering models (Collabs.) into LLM-based recommendation. LLaRA~\citep{llara} maps ID representations into behavioral tokens through a projector, and uses curriculum prompt tuning to progressively combine textual semantics with behavioral patterns. BIGRec~\citep{bigrec} does not explicitly inject collaborative embeddings into the internal representation of LLMs, but it incorporates collaborative information in a two-step grounding process and shows that LLMs still have limited ability to absorb statistical collaborative signals. CoLLM~\citep{collm} directly maps representations learned by external collaborative models into the LLM input space, making collaborative knowledge available as soft prompts. SeLLa-Rec~\citep{sellarec} further identifies the semantic gap between the collaborative space and the LLM semantic space, and alleviates this gap through semantic knowledge distillation. Beyond input-level injection, recent methods further extend this line of work to parameter and objective coupling~\citep{cora-rec,tca4rec}. However, it still pays limited attention to how collaborative embeddings are hierarchically utilized inside LLMs across Transformer depths, which is the central focus of SAILRec.

\section{Methods}

\begin{figure*}[t]
    \centering
    \includegraphics[width=1.0\textwidth]{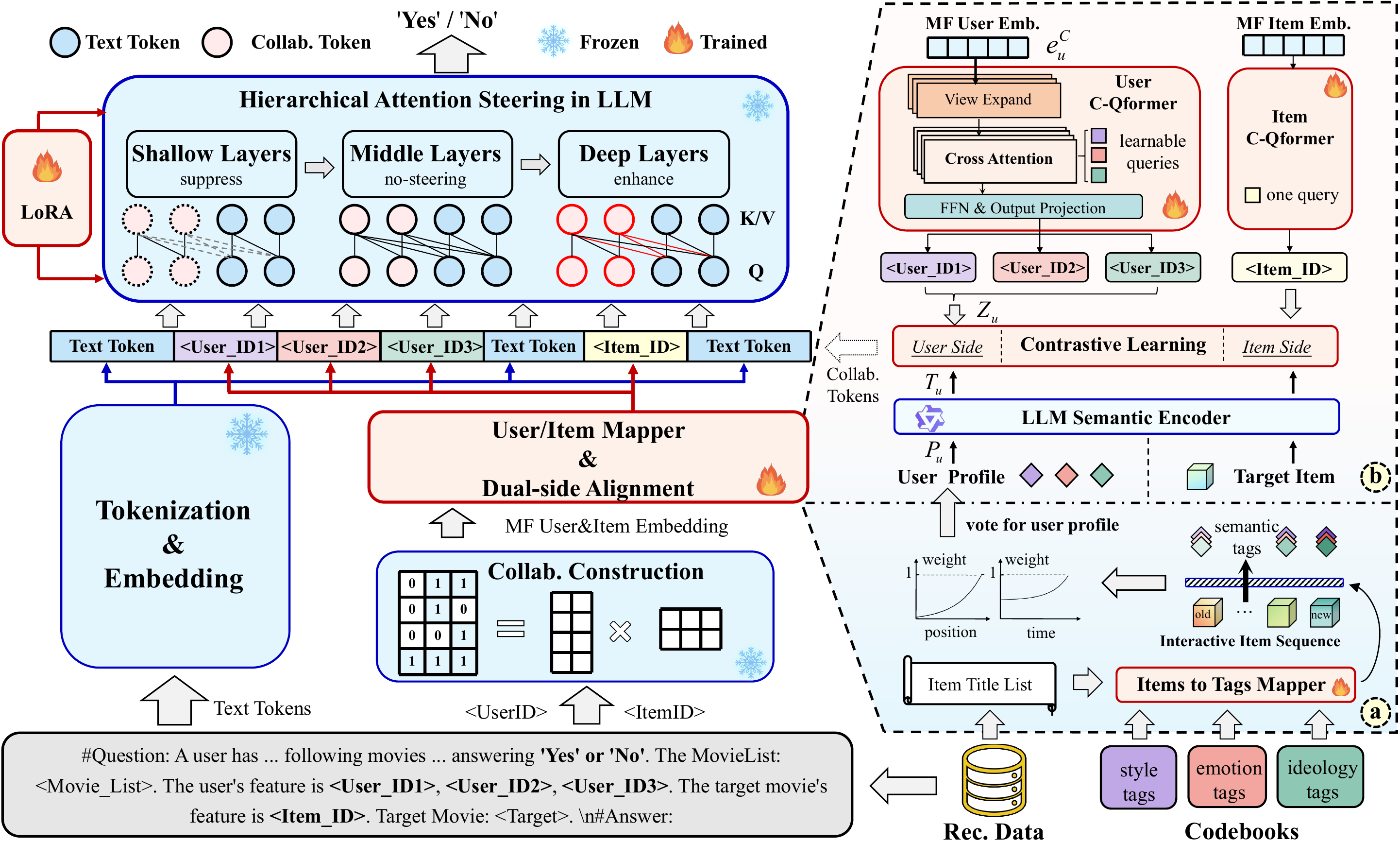}
    \caption{The overall architecture of SAILRec. The left part illustrates the main recommendation pipeline with collaborative embedding injection and hierarchical attention steering. The right part shows the dual-side warm-up process for user and item collaborative embedding mappers. Marker (a) denotes the construction of user-side alignment targets from historical interactions and semantic codebooks, while marker (b) denotes the dual-side contrastive learning process for user-side and item-side semantic alignment.}
    \label{fig:overview}
\end{figure*}
\subsection{Task Formulation}
We study click-through rate prediction in LLM-based recommendation, a standard recommendation task widely studied in conventional recommender systems~\citep{wide-deep,deepfm}. Let $\mathcal{U}$ and $\mathcal{I}$ be the user and item sets. For a user $u\in\mathcal{U}$ and a candidate item $i\in\mathcal{I}$, let $\mathcal{H}_u$ denote the historical interaction sequence and $y_{u,i}\in\{0,1\}$ denote the binary label, where $1$ indicates a click or positive preference. The task is to estimate the click probability $\hat{y}_{u,i}\in[0,1]$ according to $\mathcal{H}_u$ and $i$.

We construct a natural language prompt with the historical item list, the target item title, and two kinds of special tokens for collaborative embeddings (Collab. tokens). Prompt details are provided in Appendix~\ref{app:prompts}. At the answer position, the LLM produces vocabulary logits $\mathbf{o}_{u,i}\in\mathbb{R}^{|\mathcal{V}|}$, where $\mathcal{V}$ is the vocabulary. We take the logits of \texttt{Yes} and \texttt{No}, denoted by $o_{\text{Yes}}$ and $o_{\text{No}}$, and compute
\begin{equation}
\hat{y}_{u,i}
=
\frac{\exp(o_{\text{Yes}})}
{\exp(o_{\text{Yes}})+\exp(o_{\text{No}})}.
\label{eq:eq1}
\end{equation}
The recommendation objective is optimized with binary cross-entropy:
\begin{equation}
\begin{aligned}
\mathcal{L}_{\mathrm{rec}}
= {} & -y_{u,i}\log \hat{y}_{u,i} \\
& -(1-y_{u,i})\log(1-\hat{y}_{u,i}).
\end{aligned}
\label{eq:eq2}
\end{equation}

\subsection{Overview}
The overall architecture of SAILRec is shown in Figure~\ref{fig:overview}. Starting from recommendation data, a collaborative filtering model (Collab.) first learns user and item collaborative embeddings. These embeddings are then passed through two warm-up mapping layers and inserted into the prompt together with textual tokens. The resulting mixed embeddings are fed into the LLM, where layer-wise attention steering controls the use of collaborative information before the model predicts \texttt{`Yes'} or \texttt{`No'}. In this pipeline, dual-side semantic alignment provides reliable collaborative embeddings, and hierarchical attention steering regulates their utilization inside the LLM.

\subsection{Dual-side Alignment}
\paragraph{Collaborative filtering model (Collab.) construction.}
Before dual-side semantic alignment, SAILRec first trains a Collab. to provide task-specific collaborative embeddings for users and items. Specifically, we use matrix factorization (MF) as the collaborative model. For a user $u\in\mathcal{U}$ and an item $i\in\mathcal{I}$, MF learns a user embedding $\mathbf{e}^{c}_{u}\in\mathbb{R}^{d_c}$ and an item embedding $\mathbf{e}^{c}_{i}\in\mathbb{R}^{d_c}$, where $d_c$ denotes the dimension of collaborative embeddings. After training, the MF parameters are frozen. The learned user embedding $\mathbf{e}^{c}_{u}$ and item embedding $\mathbf{e}^{c}_{i}$ serve as the inputs for subsequent user-side and item-side semantic alignment.

\paragraph{User-side semantic alignment.}
Since anonymous user IDs have no direct linguistic meaning to the LLM, SAILRec needs structured semantic anchors to describe user preferences. Moreover, user preference is not only determined by item topics, but also by stylistic form, affective response, and high-level value orientation. Motivated by this, SAILRec derives semantic user profiles from historically interacted items with three phrase-level codebooks, $\mathcal{C}^{A}$, $\mathcal{C}^{B}$, and $\mathcal{C}^{C}$, corresponding to style, emotion, and ideology, respectively. Each codebook $\mathcal{C}^{r}=\{c^{r}_{1},\ldots,c^{r}_{K_r}\}$ contains candidate tags for semantic direction $r\in\{A,B,C\}$. We use the frozen LLM semantic encoder $E_{\mathrm{LLM}}(\cdot)$ to encode item titles and codebook tags. Examples of semantic tags in the codebooks are provided in Appendix~\ref{app:codebooks}. Then a lightweight mapper is trained to select the tag whose LLM semantic representation is most compatible with the item-title representation in each semantic direction. For user $u$ with history $\mathcal{H}_u=(i_1^u,\ldots,i_T^u)$, each historical item is weighted by positional and temporal decay. Specifically, we compute $w_j=w^{\mathrm{pos}}_j w^{\mathrm{time}}_j$ and normalize it to $\tilde{w}_j=w_j/\sum_{\ell=1}^{T}w_{\ell}$. Then, for each semantic direction, SAILRec aggregates historical tags by weighted voting and selects the top-scoring tag:
\begin{equation}
\begin{aligned}
S^{r}_{u}(c)
&=
\sum_{j=1}^{T}
\tilde{w}_{j}
\mathbb{I}\left[\phi^{r}(i^{u}_{j})=c\right], \\
p^{r}_{u}
&=
\arg\max_{c\in\mathcal{C}^{r}} S^{r}_{u}(c).
\end{aligned}
\label{eq:user_profile}
\end{equation}
The final semantic user profile is $\mathcal{P}_{u}=\{p^{A}_{u},p^{B}_{u},p^{C}_{u}\}$. The LLM semantic representations of these three profile tags are used as the alignment targets for user-side collaborative knowledge.

After constructing user-side semantic targets, SAILRec trains a user-side mapping module to align low-dimensional collaborative embeddings with the high-dimensional LLM semantic space. We design a lightweight variant of Q-Former~\citep{qformer}, named Collaborative Q-Former (C-QFormer), which uses learnable queries and cross-attention to extract collaborative information from MF embeddings. Given the frozen MF user embedding $\mathbf{e}^{c}_{u}\in\mathbb{R}^{d_c}$, the user-side C-QFormer $g_u(\cdot)$ maps it into three high-dimensional collaborative embeddings, denoted as $\mathbf{Z}_{u}=g_u(\mathbf{e}^{c}_{u})= [\mathbf{z}^{A}_{u},\mathbf{z}^{B}_{u},\mathbf{z}^{C}_{u}]\in\mathbb{R}^{3\times d_L}$, where $d_L$ is the LLM hidden size. We use special textual placeholders, namely \texttt{<User\_ID1>}, \texttt{<User\_ID2>}, and \texttt{<User\_ID3>}, to indicate the insertion positions of the three user-side collaborative embeddings. After the corresponding mapped embeddings are inserted into these positions, we refer to them as Collab. tokens for brevity.

For the semantic user profile $\mathcal{P}_{u}=\{p^{A}_{u},p^{B}_{u},p^{C}_{u}\}$, the frozen LLM semantic encoder provides target embeddings $\mathbf{T}_{u}=[\mathbf{t}^{A}_{u},\mathbf{t}^{B}_{u},\mathbf{t}^{C}_{u}]$, where $\mathbf{t}^{r}_{u}=E_{\mathrm{LLM}}(p^{r}_{u})$ and $r\in\{A,B,C\}$ corresponds to style, emotion, and ideology. We then apply slot-wise InfoNCE alignment~\citep{infonce}. For a training batch $\mathcal{B}$, the alignment loss of semantic direction $r$ is
\begin{equation}
\mathcal{L}^{r}_{u}
=
-\frac{1}{|\mathcal{B}|}
\sum_{u\in\mathcal{B}}
\log
\frac{
\exp\left(\mathrm{sim}(\mathbf{z}^{r}_{u},\mathbf{t}^{r}_{u})/\tau\right)
}{
\sum_{v\in\mathcal{B}}
\exp\left(\mathrm{sim}(\mathbf{z}^{r}_{u},\mathbf{t}^{r}_{v})/\tau\right)
},
\label{eq:eq-infonce}
\end{equation}
where $\mathrm{sim}(\cdot,\cdot)$ denotes cosine similarity and $\tau$ is the temperature. The final user-side alignment objective $\mathcal{L}_{u}$ is obtained by averaging $\mathcal{L}^{r}_{u}$ over the three semantic directions $r\in\{A,B,C\}$. Through this process, C-QFormer maps implicit user collaborative preferences into structured Collab. tokens aligned with three semantic directions, providing interpretable user-side collaborative information for LLM-based recommendation.

\paragraph{Item-side semantic alignment.}
Item-side semantic alignment is simpler because the LLM already contains world knowledge about items. We therefore use the LLM semantic representation of each item title as the alignment target. Given the frozen MF item embedding, the item-side C-QFormer uses one learnable query to map it into a high-dimensional item-side Collab. token. This token is then aligned with the corresponding item semantic representation using the InfoNCE loss, making item-side collaborative knowledge compatible with LLM item semantics.

\subsection{Hierarchical Attention Steering }
SAILRec introduces hierarchical attention steering inside the LLM by explicitly partitioning the $L$ Transformer layers into shallow, middle, and deep groups. The shallow group refers to early Transformer layers, the middle group refers to intermediate layers, and the deep group refers to top layers close to the output. Let $\mathcal{C}$ denote the key positions of user-side and item-side Collab. tokens; at layer $\ell$, SAILRec steers attention by adding a bias only to logits associated with these key positions. The attention weights are computed as
\begin{equation}
\mathbf{A}^{\ell}
=
\mathrm{softmax}\left(
\frac{\mathbf{Q}^{\ell}(\mathbf{K}^{\ell})^{\top}}{\sqrt{d_h}}
+
\mathbf{M}
+
\Delta^{\ell}
\right),
\label{eq:attn_weights}
\end{equation}
where $\mathbf{M}$ is the causal mask. The bias matrix $\Delta^{\ell}$ has the same shape as the attention logits, and its entry $\Delta^{\ell}_{q,k}$ modifies only the attention path from query position $q$ to key position $k$ when $k$ is a Collab. token position.

For shallow layers, SAILRec suppresses premature attention to Collab. tokens by assigning a negative bias to their key positions:
\begin{equation}
\Delta^{\ell}_{q,k}
=
\begin{cases}
b_s, & k\in\mathcal{C}, \\
0, & k\notin\mathcal{C},
\end{cases}
\qquad
b_s<0.
\label{eq:eq-shallow-bias}
\end{equation}
This reduces the influence of collaborative embeddings in early layers, allowing the model to focus on lexical, syntactic, and contextual understanding.

For middle layers, SAILRec keeps the original attention structure unchanged by setting $\Delta^{\ell}_{q,k}=0$ for all positions. This allows textual semantics and collaborative information to interact naturally without additional steering.

For deep layers, SAILRec strengthens the attention to Collab. tokens by assigning a positive bias to their key positions:
\begin{equation}
\Delta^{\ell}_{q,k}
=
\begin{cases}
b_d(\ell), & k\in\mathcal{C}, \\
0, & k\notin\mathcal{C},
\end{cases}
\qquad
b_d(\ell)>0.
\label{eq:eq-deep-bias}
\end{equation}
Here, $b_d(\ell)$ is a depth-dependent schedule, meaning that the positive bias changes with the layer depth. This design increases the influence of collaborative evidence near the decision stage. In this way, SAILRec regulates only the attention paths toward Collab. tokens and explicitly models their layer-wise utilization during LLM inference.

\subsection{Training Strategy}
SAILRec adopts a three-stage training strategy:

\textbf{Stage 1:} We train an MF model on recommendation data to obtain low-dimensional user and item collaborative embeddings, and then freeze it as the collaborative knowledge provider.

\textbf{Stage 2:} We respectively warm up the user-side and item-side C-QFormers with contrastive alignment, using semantic user profiles and LLM item-title representations as targets. 

\textbf{Stage 3:} We perform supervised fine-tuning (SFT) based on Low-Rank Adaptation (LoRA)~\citep{lora} for click-through rate prediction. In this stage, the MF model remains frozen, while LoRA parameters and both C-QFormers are jointly optimized under hierarchical attention steering.

\section{Experiment}

\begin{table*}[t]
\centering
\footnotesize
\setlength{\tabcolsep}{4.5pt}
\renewcommand{\arraystretch}{1.12}
\begin{tabular*}{\textwidth}{@{\extracolsep{\fill}}lcccccccc@{}}
\toprule
\multirow{2}{*}{\textbf{Method}}
& \multicolumn{4}{c}{\textbf{MovieLens-1M}}
& \multicolumn{4}{c}{\textbf{Amazon-Book}} \\
\cmidrule(lr){2-5} \cmidrule(lr){6-9}
& \textbf{AUC} & \textbf{UAUC} & \textbf{NDCG} & \textbf{MAP}
& \textbf{AUC} & \textbf{UAUC} & \textbf{NDCG} & \textbf{MAP} \\
\midrule

\rowcolor{gray!12}
\multicolumn{9}{@{}l}{\textit{Collaborative filtering models}} \\
MF
& 0.6429 & 0.6126 & 0.8422 & 0.7093
& 0.7098 & 0.5630 & 0.8221 & 0.7420 \\
LightGCN
& 0.5831 & 0.6493 & 0.8532 & 0.7320
& 0.7001 & 0.5608 & 0.8218 & 0.7415 \\
SASRec
& 0.7010 & 0.6742 & 0.8581 & 0.7411
& 0.6646 & 0.5629 & 0.8208 & 0.7393 \\

\midrule
\rowcolor{gray!12}
\multicolumn{9}{@{}l}{\textit{LLM-only recommenders}} \\
TALLRec
& 0.7039 & 0.6837 & 0.8701 & 0.7598
& 0.7198 & 0.6524 & 0.8624 & 0.7959 \\
GDRT
& 0.7249 & 0.6871 & 0.8689 & 0.7537
& 0.7607 & 0.6280 & 0.8505 & 0.7800 \\
HatLLM
& 0.7330 & 0.7058 & 0.8762 & 0.7645
& 0.7397 & 0.6147 & 0.8475 & 0.7754 \\

\midrule
\rowcolor{gray!12}
\multicolumn{9}{@{}l}{\textit{Collaborative-enhanced LLM recommenders}} \\
CoLLM
& 0.7137 & 0.6850 & 0.8757 & 0.7647
& 0.7924 & 0.7197 & 0.8896 & 0.8351 \\
BinLLM
& 0.7271 & 0.6980 & 0.8833 & 0.7779
& 0.8031 & 0.7234 & 0.8922 & 0.8384 \\
CoRA
& 0.7284 & 0.6956 & 0.8744 & 0.7614
& 0.8156 & 0.7201 & 0.8930 & 0.8393 \\
CKF
& 0.7320 & 0.7120 & 0.8884 & \underline{0.7830}
& 0.8048 & 0.7246 & 0.8929 & 0.8398 \\
TokenRec
& 0.7272 & 0.7036 & 0.8798 & 0.7702
& 0.7975 & 0.6027 & 0.8360 & 0.7603 \\
SeLLa-Rec
& \underline{0.7428} & \underline{0.7133} & \underline{0.8885} & 0.7825
& \underline{0.8191} & \underline{0.7310} & \underline{0.8940} & \underline{0.8419} \\
TCA4Rec
& 0.7353 & 0.7123 & 0.8838 & 0.7733
& 0.7703 & 0.6208 & 0.8464 & 0.7747 \\
\textbf{Ours (SAILRec)}
& \textbf{0.7571} & \textbf{0.7410} & \textbf{0.8985} & \textbf{0.7999}
& \textbf{0.8393} & \textbf{0.7506} & \textbf{0.9017} & \textbf{0.8524} \\
\bottomrule
\end{tabular*}
\caption{Overall recommendation performance on MovieLens-1M and Amazon-Book. The best results are in bold, and the second-best results are underlined.}
\label{tab:main_results}
\end{table*}

\subsection{Experimental Settings}
\paragraph{Datasets.}
Following the experimental settings and data processing protocols of CoLLM~\citep{collm} and SeLLa-Rec~\citep{sellarec}, we evaluate SAILRec on two public datasets, MovieLens-1M~\citep{movielens-1m} and Amazon-Book~\citep{amazon-book}. MovieLens-1M is suitable for evaluating recommendation performance in a relatively dense movie preference scenario, while Amazon-Book contains more diverse user interests, stronger long-tail item distributions, and sparser interactions. Detailed data processing procedures are provided in Appendix~\ref{app:datasets}.

\paragraph{Baselines.}
We compare SAILRec with three groups of representative baselines. The first group includes collaborative filtering models (Collabs.), namely MF~\citep{mf}, LightGCN~\citep{lightgcn}, and SASRec~\citep{sasrec}, which learn user and item representations directly from interaction data. The second group includes LLM-only recommenders, namely TALLRec~\citep{tallrec}, GDRT~\citep{gdrt}, and HatLLM~\citep{hatllm}. These methods rely on the semantic understanding, instruction following, and attention modeling abilities of LLMs for preference prediction. The third group includes collaborative-enhanced LLM recommenders, including CoLLM~\citep{collm}, BinLLM~\citep{binllm}, CoRA~\citep{cora-rec}, CKF~\citep{CKF}, TokenRec~\citep{tokenrec}, SeLLa-Rec~\citep{sellarec}, and TCA4Rec~\citep{tca4rec}. These methods inject collaborative knowledge learned from collaborative models or ID representations into LLMs to compensate for the limited modeling of user IDs and item IDs. Detailed descriptions of the baselines are provided in Appendix~\ref{app:baselines}.

\paragraph{Metrics.}
We evaluate model performance using \textit{Area Under the ROC Curve} (AUC)~\citep{auc}, \textit{User-level Area Under the ROC Curve} (UAUC), \textit{Normalized Discounted Cumulative Gain} (NDCG)~\citep{ndcg}, and \textit{Mean Average Precision} (MAP). These metrics can measure model performance from different perspectives~\citep{why-4metrics,survey-rec-metrics,use-map}. Detailed implementation is provided in Appendix~\ref{app:metrics}.

\paragraph{Other Settings.}
We use MF as the collaborative model, with the collaborative embedding dimension set to 256. The LLM backbone is Qwen2-7B~\citep{qwen2}, whose hidden size is 3584. For hierarchical attention steering, SAILRec sets the shallow-layer ratio to 30\%, the deep-layer ratio to 20\%, the shallow-layer bias to $-1.0$, the maximum deep-layer bias to $2.0$, and uses a triangular schedule for deep-layer bias. More details are provided in Appendix~\ref{app:other_settings}.

\subsection{Overall Performance}

Table~\ref{tab:main_results} reports the overall performance. SAILRec achieves the best results on both datasets across all metrics. On MovieLens-1M, it improves over the strongest baseline by 1.43, 2.77, 1.00, and 1.69 percentage points on AUC, UAUC, NDCG, and MAP, respectively. On Amazon-Book, the corresponding gains are 2.02, 1.96, 0.77, and 1.05 percentage points. These results show that SAILRec consistently outperforms conventional collaborative filtering models, LLM-only recommenders, and collaborative-enhanced LLM recommenders. The larger gains on UAUC and MAP indicate that SAILRec improves not only global positive-negative discrimination, but also user-level preference modeling and ranking quality. The consistent improvements on both MovieLens-1M and Amazon-Book further suggest that the method is effective in both relatively dense movie recommendation and sparser book recommendation scenarios. Compared with CoLLM, which directly injects unaligned collaborative embeddings, and SeLLa-Rec, which mainly performs item-side alignment, SAILRec achieves stronger results, suggesting that collaborative knowledge should be made semantically accessible on both user and item sides. Compared with HatLLM, which steers attention over textual interaction sequences in an LLM-only setting, SAILRec further shows that attention control is also effective for injected collaborative embeddings when combined with semantic alignment. Overall, the gains indicate that dual-side semantic alignment and hierarchical attention steering jointly enable more effective use of collaborative information inside LLMs.

\begin{figure}[t]
    \centering
    \includegraphics[width=1\columnwidth]{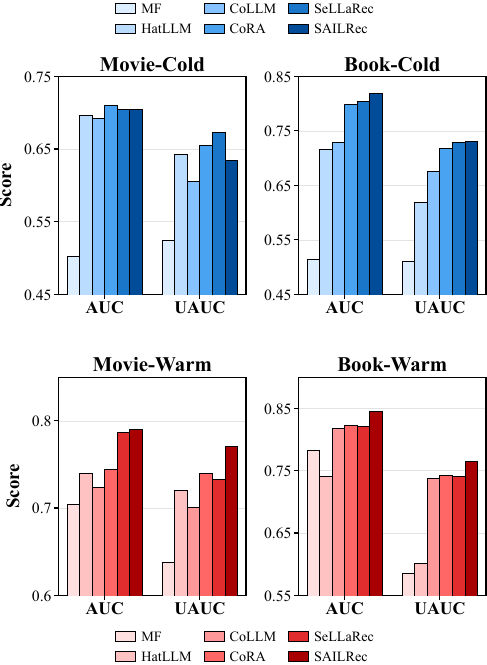}
    \caption{Models’ performance on warm/cold data.}
    \label{fig:warm-cold}
\end{figure}

\begin{table}[H]
\centering
\scriptsize
\setlength{\tabcolsep}{3.2pt}
\renewcommand{\arraystretch}{1.08}
\resizebox{\columnwidth}{!}{
\begin{tabular}{@{}lcccccc@{}}
\toprule
\multirow{2}{*}{\textbf{Model}}
& \multicolumn{3}{c}{\textbf{MovieLens-1M}}
& \multicolumn{3}{c}{\textbf{Amazon-Book}} \\
\cmidrule(lr){2-4} \cmidrule(lr){5-7}
& \textbf{AUC} & \textbf{UAUC} & \textbf{NDCG}
& \textbf{AUC} & \textbf{UAUC} & \textbf{NDCG} \\
\midrule

\rowcolor{gray!12}
\multicolumn{7}{@{}l}{\textit{Model Structure Variants}} \\
SAIL-w/o-S
& 0.7546 & 0.7268 & 0.8923
& 0.8220 & 0.7194 & 0.8896 \\
SAIL-w/o-UI
& 0.7322 & 0.7236 & 0.8894
& 0.8116 & 0.7391 & 0.8989 \\
SAIL-w/o-I
& 0.7346 & 0.7277 & 0.8971
& 0.8267 & 0.7441 & 0.8991 \\
SAIL-w/o-U
& 0.7434 & 0.7205 & 0.8932
& 0.8288 & 0.7369 & 0.8980 \\
SAIL-w/o-C
& 0.7475 & 0.7295 & 0.8953
& 0.8310 & 0.7367 & 0.8967 \\

\midrule
\rowcolor{gray!12}
\multicolumn{7}{@{}l}{\textit{Tuning Method Variants}} \\
SAIL-LoRA
& 0.7283 & 0.7116 & 0.8872
& 0.7649 & 0.6928 & 0.8812 \\
SAIL-LoRA-Map
& 0.7507 & 0.7225 & 0.8980
& 0.8301 & 0.7375 & 0.8983 \\

\midrule
Full-SAILRec
& \textbf{0.7571} & \textbf{0.7410} & \textbf{0.8985}
& \textbf{0.8393} & \textbf{0.7506} & \textbf{0.9017} \\
\bottomrule
\end{tabular}
}
\caption{Performance comparison for SAILRec and variants. Here, S, U, I, and C denote attention steering, user-side alignment, item-side alignment, and codebooks. }
\label{tab:ablation}
\end{table}

\subsection{Performance on Warm/Cold Data}

Figure~\ref{fig:warm-cold} compares representative methods on warm and cold subsets. SAILRec shows stronger advantages on warm data, where richer interaction histories make collaborative embeddings more reliable and allow dual-side alignment and attention steering to better exploit collaborative evidence. On cold data, SAILRec remains competitive, especially on Book-Cold, while the gains on Movie-Cold are smaller. This suggests that cold-start prediction still relies heavily on textual semantics and that collaborative enhancement should be carefully controlled rather than uniformly amplified.

\begin{table*}[t]
\centering
\footnotesize
\setlength{\tabcolsep}{5.0pt}
\renewcommand{\arraystretch}{1.12}
\begin{tabular*}{\textwidth}{@{\extracolsep{\fill}}lcccccccc@{}}
\toprule
\multirow{2}{*}{\textbf{Method}}
& \multicolumn{4}{c}{\textbf{MovieLens-1M}}
& \multicolumn{4}{c}{\textbf{Amazon-Book}} \\
\cmidrule(lr){2-5} \cmidrule(lr){6-9}
& \textbf{AUC} & \textbf{UAUC} & \textbf{NDCG} & \textbf{MAP}
& \textbf{AUC} & \textbf{UAUC} & \textbf{NDCG} & \textbf{MAP} \\
\midrule
\textbf{SAIL-S-N-E}
& \textbf{0.7571} & \textbf{0.7410} & 0.8985 & \textbf{0.7999}
& 0.8393 & \textbf{0.7506} & \textbf{0.9017} & \textbf{0.8524} \\
SAIL-S-E-N
& 0.7487 & 0.7377 & 0.8948 & 0.7943
& 0.8344 & 0.7401 & 0.8983 & 0.8477 \\
SAIL-E-S-N
& 0.7480 & 0.7238 & 0.8949 & 0.7933
& 0.8367 & 0.7418 & 0.8986 & 0.8482 \\
SAIL-E-N-S
& 0.7461 & 0.7352 & \textbf{0.9008} & 0.7996
& \textbf{0.8399} & 0.7415 & 0.8992 & 0.8490 \\
SAIL-N-E-S
& 0.7484 & 0.7335 & 0.8939 & 0.7932
& 0.8324 & 0.7386 & 0.8984 & 0.8476 \\
SAIL-N-S-E
& 0.7505 & 0.7266 & 0.8954 & 0.7943
& 0.8357 & 0.7307 & 0.8948 & 0.8434 \\
\midrule
SAIL-S
& 0.7503 & 0.7274 & 0.8963 & 0.7966
& 0.8281 & 0.7449 & 0.9000 & 0.8520 \\
SAIL-N
& 0.7546 & 0.7268 & 0.8923 & 0.7894
& 0.8220 & 0.7194 & 0.8896 & 0.8359 \\
SAIL-E
& 0.7449 & 0.7351 & 0.8965 & 0.7974
& 0.8290 & 0.7447 & 0.9008 & 0.8510 \\
\bottomrule
\end{tabular*}
\caption{Performance comparison of different layer-wise attention steering schedules. S, N, and E denote suppression, no steering, and enhancement, respectively.}
\label{tab:attention_strategy}
\end{table*}

\subsection{Ablation Study}
Table~\ref{tab:ablation} reports the ablation results of SAILRec. SAIL-w/o-S removes the hierarchical attention steering mechanism. Its performance drop shows that aligned collaborative embeddings are not sufficient by themselves, since the LLM still benefits from explicit layer-wise control over when collaborative evidence should be emphasized. SAIL-w/o-UI removes semantic alignment on both user and item sides, while SAIL-w/o-U and SAIL-w/o-I remove user-side and item-side alignment, respectively. The degradation of these variants indicates that collaborative embeddings need to be made semantically compatible with the LLM before being injected into the prompt. The weaker performance of SAIL-w/o-U further shows the importance of user-side alignment, because anonymous user IDs do not provide linguistic cues and require explicit preference anchors for grounding. SAIL-w/o-C removes the codebook-based user profiling strategy and instead uses the embedding encoded from the whole historical interaction sequence as the user-side alignment reference. Its lower performance suggests that a single sequence-level representation is less effective than structured user profiles built from semantic tags, as the codebook design offers clearer and more stable anchors for aligning user-side collaborative preferences.

For training variants, SAIL-LoRA trains only the LoRA parameters in the final stage and freezes both mapping modules, while SAIL-LoRA-Map further updates the mapping modules but does not jointly optimize them with LoRA from the beginning of the final recommendation stage. Their weaker results indicate that collaborative mappings should be adapted together with the LLM to fit the recommendation objective.

\subsection{Steering Strategy Analysis}
\label{sec:steering_strategy_analysis}

To further examine how collaborative embeddings should be used across Transformer depths, we compare different layer-wise attention steering schedules in Table~\ref{tab:attention_strategy}. We divide the Transformer layers into shallow, middle, and deep groups, which account for 30\%, 50\%, and 20\% of all layers, respectively, and then test different steering operations on these groups. Here, S, N, and E denote suppression, no steering, and enhancement, respectively. For example, SAIL-S-N-E applies suppression in shallow layers, no additional steering in middle layers, and enhancement in deep layers. The single-strategy variants SAIL-S, SAIL-N, and SAIL-E apply the same steering operation to all layers.

The results show that the default SAIL-S-N-E schedule achieves the most consistent performance across the two datasets. Although SAIL-E-N-S obtains slightly higher NDCG on MovieLens-1M and AUC on Amazon-Book, its UAUC is less stable, suggesting that steering strategies should consider both user-level discrimination and ranking quality. The comparison supports the layer-wise utilization pattern in our diagnostic analysis. Schedules with shallow-layer suppression generally outperform those that enhance or leave collaborative embeddings uncontrolled in early layers, indicating that collaborative knowledge should not interfere too early with textual encoding. Comparing SAIL-S-N-E with SAIL-S-E-N shows that deep-layer enhancement is more effective than middle-layer enhancement, suggesting that collaborative evidence is more useful near the decision stage. The weaker results of SAIL-S, SAIL-N, and SAIL-E show that uniform steering is insufficient. Overall, effective collaborative enhancement requires selective layer-wise access rather than uniform attention adjustment. Appendix~\ref{app:attention_bias_masking} and Appendix~\ref{app:qwen3_results} further provide masking and Qwen3-8B backbone evidence supporting the effectiveness of SAILRec.

\section{Conclusion}

In this paper, we present SAILRec, an LLM-based recommender that improves the use of collaborative embeddings through dual-side semantic alignment and hierarchical attention steering. SAILRec aligns user-side and item-side collaborative signals with the LLM semantic space and regulates their influence across Transformer layers to better balance semantic and collaborative knowledge. Experiments on MovieLens-1M and Amazon-Book show consistent improvements over representative baselines, while ablations and masking analyses validate the proposed designs.

\section*{Limitations}

This work has two main limitations. First, SAILRec uses a predefined attention steering schedule, which is effective in our experiments but may not be optimal for all samples or LLM backbones. Future work may develop sample-adaptive steering with continuous layer-wise bias functions, allowing different samples to receive customized Collab. token biases across Transformer layers. Second, although our experiments provide direct evidence that Collab. tokens affect prediction, their semantic interpretability remains limited. Since these tokens are still continuous embeddings, they cannot be directly interpreted as explicit natural language concepts. More interpretable collaborative tokenization and semantic grounding are useful future directions.


\bibliography{custom}

@misc{qwen3,
      title={Qwen3 Technical Report}, 
      author={An Yang and Anfeng Li and Baosong Yang and Beichen Zhang and Binyuan Hui and Bo Zheng and Bowen Yu and Chang Gao and Chengen Huang and Chenxu Lv and Chujie Zheng and Dayiheng Liu and Fan Zhou and Fei Huang and Feng Hu and Hao Ge and Haoran Wei and Huan Lin and Jialong Tang and Jian Yang and Jianhong Tu and Jianwei Zhang and Jianxin Yang and Jiaxi Yang and Jing Zhou and Jingren Zhou and Junyang Lin and Kai Dang and Keqin Bao and Kexin Yang and Le Yu and Lianghao Deng and Mei Li and Mingfeng Xue and Mingze Li and Pei Zhang and Peng Wang and Qin Zhu and Rui Men and Ruize Gao and Shixuan Liu and Shuang Luo and Tianhao Li and Tianyi Tang and Wenbiao Yin and Xingzhang Ren and Xinyu Wang and Xinyu Zhang and Xuancheng Ren and Yang Fan and Yang Su and Yichang Zhang and Yinger Zhang and Yu Wan and Yuqiong Liu and Zekun Wang and Zeyu Cui and Zhenru Zhang and Zhipeng Zhou and Zihan Qiu},
      year={2025},
      eprint={2505.09388},
      archivePrefix={arXiv},
      primaryClass={cs.CL},
      url={https://arxiv.org/abs/2505.09388}, 
}

@inproceedings{bert-pipeline,
  title={BERT Rediscovers the Classical NLP Pipeline},
  author={Ian Tenney and Dipanjan Das and Ellie Pavlick},
  booktitle={Annual Meeting of the Association for Computational Linguistics},
  year={2019},
  url={https://api.semanticscholar.org/CorpusID:155092004}
}

@inproceedings{quantify-attention,
    title = "Quantifying Attention Flow in Transformers",
    author = "Abnar, Samira  and
      Zuidema, Willem",
    editor = "Jurafsky, Dan  and
      Chai, Joyce  and
      Schluter, Natalie  and
      Tetreault, Joel",
    booktitle = "Proceedings of the 58th Annual Meeting of the Association for Computational Linguistics",
    month = jul,
    year = "2020",
    address = "Online",
    publisher = "Association for Computational Linguistics",
    url = "https://aclanthology.org/2020.acl-main.385/",
    doi = "10.18653/v1/2020.acl-main.385",
    pages = "4190--4197"
}

@inproceedings{bert-attention,
    title = "What Does {BERT} Look at? An Analysis of {BERT}{'}s Attention",
    author = "Clark, Kevin  and
      Khandelwal, Urvashi  and
      Levy, Omer  and
      Manning, Christopher D.",
    editor = "Linzen, Tal  and
      Chrupa{\l}a, Grzegorz  and
      Belinkov, Yonatan  and
      Hupkes, Dieuwke",
    booktitle = "Proceedings of the 2019 ACL Workshop BlackboxNLP: Analyzing and Interpreting Neural Networks for NLP",
    month = aug,
    year = "2019",
    address = "Florence, Italy",
    publisher = "Association for Computational Linguistics",
    url = "https://aclanthology.org/W19-4828/",
    doi = "10.18653/v1/W19-4828",
    pages = "276--286"
}

@inproceedings{attention,
author = {Vaswani, Ashish and Shazeer, Noam and Parmar, Niki and Uszkoreit, Jakob and Jones, Llion and Gomez, Aidan N. and Kaiser, \L{}ukasz and Polosukhin, Illia},
title = {Attention is all you need},
year = {2017},
isbn = {9781510860964},
publisher = {Curran Associates Inc.},
address = {Red Hook, NY, USA},
booktitle = {Proceedings of the 31st International Conference on Neural Information Processing Systems},
pages = {6000–6010},
numpages = {11},
location = {Long Beach, California, USA},
series = {NIPS'17}
}

@inproceedings{qformer,
author = {Li, Junnan and Li, Dongxu and Savarese, Silvio and Hoi, Steven},
title = {BLIP-2: bootstrapping language-image pre-training with frozen image encoders and large language models},
year = {2023},
publisher = {JMLR.org},
booktitle = {Proceedings of the 40th International Conference on Machine Learning},
articleno = {814},
numpages = {13},
location = {Honolulu, Hawaii, USA},
series = {ICML'23}
}

@inproceedings{p5,
author = {Geng, Shijie and Liu, Shuchang and Fu, Zuohui and Ge, Yingqiang and Zhang, Yongfeng},
title = {Recommendation as Language Processing (RLP): A Unified Pretrain, Personalized Prompt \& Predict Paradigm (P5)},
year = {2022},
isbn = {9781450392785},
publisher = {Association for Computing Machinery},
address = {New York, NY, USA},
url = {https://doi.org/10.1145/3523227.3546767},
doi = {10.1145/3523227.3546767},
booktitle = {Proceedings of the 16th ACM Conference on Recommender Systems},
pages = {299–315},
numpages = {17},
location = {Seattle, WA, USA},
series = {RecSys '22}
}

@inproceedings{wide-deep,
author = {Cheng, Heng-Tze and Koc, Levent and Harmsen, Jeremiah and Shaked, Tal and Chandra, Tushar and Aradhye, Hrishi and Anderson, Glen and Corrado, Greg and Chai, Wei and Ispir, Mustafa and Anil, Rohan and Haque, Zakaria and Hong, Lichan and Jain, Vihan and Liu, Xiaobing and Shah, Hemal},
title = {Wide \& Deep Learning for Recommender Systems},
year = {2016},
isbn = {9781450347952},
publisher = {Association for Computing Machinery},
address = {New York, NY, USA},
url = {https://doi.org/10.1145/2988450.2988454},
doi = {10.1145/2988450.2988454},
booktitle = {Proceedings of the 1st Workshop on Deep Learning for Recommender Systems},
pages = {7–10},
numpages = {4},
location = {Boston, MA, USA},
series = {DLRS 2016}
}

@inproceedings{deepfm,
author = {Guo, Huifeng and Tang, Ruiming and Ye, Yunming and Li, Zhenguo and He, Xiuqiang},
title = {DeepFM: a factorization-machine based neural network for CTR prediction},
year = {2017},
isbn = {9780999241103},
publisher = {AAAI Press},
booktitle = {Proceedings of the 26th International Joint Conference on Artificial Intelligence},
pages = {1725–1731},
numpages = {7},
location = {Melbourne, Australia},
series = {IJCAI'17}
}

@misc{infonce,
      title={Representation Learning with Contrastive Predictive Coding}, 
      author={Aaron van den Oord and Yazhe Li and Oriol Vinyals},
      year={2019},
      eprint={1807.03748},
      archivePrefix={arXiv},
      primaryClass={cs.LG},
      url={https://arxiv.org/abs/1807.03748}, 
}

@misc{survey-llmrec-2024-1,
      title={Recommendation with Generative Models}, 
      author={Yashar Deldjoo and Zhankui He and Julian McAuley and Anton Korikov and Scott Sanner and Arnau Ramisa and Rene Vidal and Maheswaran Sathiamoorthy and Atoosa Kasrizadeh and Silvia Milano and Francesco Ricci},
      year={2024},
      eprint={2409.15173},
      archivePrefix={arXiv},
      primaryClass={cs.IR},
      url={https://arxiv.org/abs/2409.15173}, 
}

@article{survey-llmrec-2024-2,
author = {Wu, Likang and Zheng, Zhi and Qiu, Zhaopeng and Wang, Hao and Gu, Hongchao and Shen, Tingjia and Qin, Chuan and Zhu, Chen and Zhu, Hengshu and Liu, Qi and Xiong, Hui and Chen, Enhong},
title = {A survey on large language models for recommendation},
year = {2024},
issue_date = {Sep 2024},
publisher = {Kluwer Academic Publishers},
address = {USA},
volume = {27},
number = {5},
issn = {1386-145X},
url = {https://doi.org/10.1007/s11280-024-01291-2},
doi = {10.1007/s11280-024-01291-2},
journal = {World Wide Web},
month = aug,
numpages = {31}
}

@ARTICLE{survey-rec-metrics,
  author={Wu, Le and He, Xiangnan and Wang, Xiang and Zhang, Kun and Wang, Meng},
  journal={IEEE Transactions on Knowledge and Data Engineering}, 
  title={A Survey on Accuracy-Oriented Neural Recommendation: From Collaborative Filtering to Information-Rich Recommendation}, 
  year={2023},
  volume={35},
  number={5},
  pages={4425-4445},
  doi={10.1109/TKDE.2022.3145690}}

@misc{survey-llmrec-2023-1,
      title={Large Language Models for Generative Recommendation: A Survey and Visionary Discussions}, 
      author={Lei Li and Yongfeng Zhang and Dugang Liu and Li Chen},
      year={2024},
      eprint={2309.01157},
      archivePrefix={arXiv},
      primaryClass={cs.IR},
      url={https://arxiv.org/abs/2309.01157}, 
}

@inproceedings{rec-lm,
    title = "{R}ec{LM}: Recommendation Instruction Tuning",
    author = "Jiang, Yangqin  and
      Yang, Yuhao  and
      Xia, Lianghao  and
      Luo, Da  and
      Lin, Kangyi  and
      Huang, Chao",
    editor = "Che, Wanxiang  and
      Nabende, Joyce  and
      Shutova, Ekaterina  and
      Pilehvar, Mohammad Taher",
    booktitle = "Proceedings of the 63rd Annual Meeting of the Association for Computational Linguistics (Volume 1: Long Papers)",
    month = jul,
    year = "2025",
    address = "Vienna, Austria",
    publisher = "Association for Computational Linguistics",
    url = "https://aclanthology.org/2025.acl-long.751/",
    doi = "10.18653/v1/2025.acl-long.751",
    pages = "15443--15459",
    ISBN = "979-8-89176-251-0"
}

@article{rec-ranker,
author = {Luo, Sichun and He, Bowei and Zhao, Haohan and Shao, Wei and Qi, Yanlin and Huang, Yinya and Zhou, Aojun and Yao, Yuxuan and Li, Zongpeng and Xiao, Yuanzhang and Zhan, Mingjie and Song, Linqi},
title = {RecRanker: Instruction Tuning Large Language Model as Ranker for Top-k Recommendation},
year = {2025},
issue_date = {September 2025},
publisher = {Association for Computing Machinery},
address = {New York, NY, USA},
volume = {43},
number = {5},
issn = {1046-8188},
url = {https://doi.org/10.1145/3705728},
doi = {10.1145/3705728},
journal = {ACM Trans. Inf. Syst.},
month = jul,
articleno = {113},
numpages = {31}
}

@inproceedings{llara,
author = {Liao, Jiayi and Li, Sihang and Yang, Zhengyi and Wu, Jiancan and Yuan, Yancheng and Wang, Xiang and He, Xiangnan},
title = {LLaRA: Large Language-Recommendation Assistant},
year = {2024},
isbn = {9798400704314},
publisher = {Association for Computing Machinery},
address = {New York, NY, USA},
url = {https://doi.org/10.1145/3626772.3657690},
doi = {10.1145/3626772.3657690},
booktitle = {Proceedings of the 47th International ACM SIGIR Conference on Research and Development in Information Retrieval},
pages = {1785–1795},
numpages = {11},
location = {Washington DC, USA},
series = {SIGIR '24}
}

@inproceedings{amazon-book,
    title = "Justifying Recommendations using Distantly-Labeled Reviews and Fine-Grained Aspects",
    author = "Ni, Jianmo  and
      Li, Jiacheng  and
      McAuley, Julian",
    editor = "Inui, Kentaro  and
      Jiang, Jing  and
      Ng, Vincent  and
      Wan, Xiaojun",
    booktitle = "Proceedings of the 2019 Conference on Empirical Methods in Natural Language Processing and the 9th International Joint Conference on Natural Language Processing (EMNLP-IJCNLP)",
    month = nov,
    year = "2019",
    address = "Hong Kong, China",
    publisher = "Association for Computational Linguistics",
    url = "https://aclanthology.org/D19-1018/",
    doi = "10.18653/v1/D19-1018",
    pages = "188--197"
}

@misc{qwen2,
      title={Qwen2 Technical Report}, 
      author={An Yang and Baosong Yang and Binyuan Hui and Bo Zheng and Bowen Yu and Chang Zhou and Chengpeng Li and Chengyuan Li and Dayiheng Liu and Fei Huang and Guanting Dong and Haoran Wei and Huan Lin and Jialong Tang and Jialin Wang and Jian Yang and Jianhong Tu and Jianwei Zhang and Jianxin Ma and Jianxin Yang and Jin Xu and Jingren Zhou and Jinze Bai and Jinzheng He and Junyang Lin and Kai Dang and Keming Lu and Keqin Chen and Kexin Yang and Mei Li and Mingfeng Xue and Na Ni and Pei Zhang and Peng Wang and Ru Peng and Rui Men and Ruize Gao and Runji Lin and Shijie Wang and Shuai Bai and Sinan Tan and Tianhang Zhu and Tianhao Li and Tianyu Liu and Wenbin Ge and Xiaodong Deng and Xiaohuan Zhou and Xingzhang Ren and Xinyu Zhang and Xipin Wei and Xuancheng Ren and Xuejing Liu and Yang Fan and Yang Yao and Yichang Zhang and Yu Wan and Yunfei Chu and Yuqiong Liu and Zeyu Cui and Zhenru Zhang and Zhifang Guo and Zhihao Fan},
      year={2024},
      eprint={2407.10671},
      archivePrefix={arXiv},
      primaryClass={cs.CL},
      url={https://arxiv.org/abs/2407.10671}, 
}

@article{bigrec,
author = {Bao, Keqin and Zhang, Jizhi and Wang, Wenjie and Zhang, Yang and Yang, Zhengyi and Luo, Yanchen and Chen, Chong and Feng, Fuli and Tian, Qi},
title = {A Bi-Step Grounding Paradigm for Large Language Models in Recommendation Systems},
year = {2025},
issue_date = {December 2025},
publisher = {Association for Computing Machinery},
address = {New York, NY, USA},
volume = {3},
number = {4},
url = {https://doi.org/10.1145/3716393},
doi = {10.1145/3716393},
journal = {ACM Trans. Recomm. Syst.},
month = apr,
articleno = {53},
numpages = {27}
}

@inproceedings{tallrec,
author = {Bao, Keqin and Zhang, Jizhi and Zhang, Yang and Wang, Wenjie and Feng, Fuli and He, Xiangnan},
title = {TALLRec: An Effective and Efficient Tuning Framework to Align Large Language Model with Recommendation},
year = {2023},
isbn = {9798400702419},
publisher = {Association for Computing Machinery},
address = {New York, NY, USA},
url = {https://doi.org/10.1145/3604915.3608857},
doi = {10.1145/3604915.3608857},
booktitle = {Proceedings of the 17th ACM Conference on Recommender Systems},
pages = {1007–1014},
numpages = {8},
location = {Singapore, Singapore},
series = {RecSys '23}
}

@inproceedings{why-4metrics,
author = {Davis, Jesse and Goadrich, Mark},
title = {The relationship between Precision-Recall and ROC curves},
year = {2006},
isbn = {1595933832},
publisher = {Association for Computing Machinery},
address = {New York, NY, USA},
url = {https://doi.org/10.1145/1143844.1143874},
doi = {10.1145/1143844.1143874},
booktitle = {Proceedings of the 23rd International Conference on Machine Learning},
pages = {233–240},
numpages = {8},
location = {Pittsburgh, Pennsylvania, USA},
series = {ICML '06}
}

@misc{chatrec,
      title={Chat-REC: Towards Interactive and Explainable LLMs-Augmented Recommender System}, 
      author={Yunfan Gao and Tao Sheng and Youlin Xiang and Yun Xiong and Haofen Wang and Jiawei Zhang},
      year={2023},
      eprint={2303.14524},
      archivePrefix={arXiv},
      primaryClass={cs.IR},
      url={https://arxiv.org/abs/2303.14524}, 
}

@article{movielens-1m,
author = {Harper, F. Maxwell and Konstan, Joseph A.},
title = {The MovieLens Datasets: History and Context},
year = {2015},
issue_date = {January 2016},
publisher = {Association for Computing Machinery},
address = {New York, NY, USA},
volume = {5},
number = {4},
issn = {2160-6455},
url = {https://doi.org/10.1145/2827872},
doi = {10.1145/2827872},
journal = {ACM Trans. Interact. Intell. Syst.},
month = dec,
articleno = {19},
numpages = {19}
}

@inproceedings{lightgcn,
author = {He, Xiangnan and Deng, Kuan and Wang, Xiang and Li, Yan and Zhang, YongDong and Wang, Meng},
title = {LightGCN: Simplifying and Powering Graph Convolution Network for Recommendation},
year = {2020},
isbn = {9781450380164},
publisher = {Association for Computing Machinery},
address = {New York, NY, USA},
url = {https://doi.org/10.1145/3397271.3401063},
doi = {10.1145/3397271.3401063},
booktitle = {Proceedings of the 43rd International ACM SIGIR Conference on Research and Development in Information Retrieval},
pages = {639–648},
numpages = {10},
location = {Virtual Event, China},
series = {SIGIR '20}
}

@inproceedings{
lora,
title={Lo{RA}: Low-Rank Adaptation of Large Language Models},
author={Edward J Hu and yelong shen and Phillip Wallis and Zeyuan Allen-Zhu and Yuanzhi Li and Shean Wang and Lu Wang and Weizhu Chen},
booktitle={International Conference on Learning Representations},
year={2022},
url={https://openreview.net/forum?id=nZeVKeeFYf9}
}

@INPROCEEDINGS {sasrec,
author = { Kang, Wang-Cheng and McAuley, Julian },
booktitle = { 2018 IEEE International Conference on Data Mining (ICDM) },
title = {{ Self-Attentive Sequential Recommendation }},
year = {2018},
volume = {},
ISSN = {},
pages = {197-206},
doi = {10.1109/ICDM.2018.00035},
url = {https://doi.ieeecomputersociety.org/10.1109/ICDM.2018.00035},
publisher = {IEEE Computer Society},
address = {Los Alamitos, CA, USA},
month =Nov}

@ARTICLE{mf,
  author={Koren, Yehuda and Bell, Robert and Volinsky, Chris},
  journal={Computer}, 
  title={Matrix Factorization Techniques for Recommender Systems}, 
  year={2009},
  volume={42},
  number={8},
  pages={30-37},
  doi={10.1109/MC.2009.263}}

@article{auc,
title = {An introduction to ROC analysis},
journal = {Pattern Recognition Letters},
volume = {27},
number = {8},
pages = {861-874},
year = {2006},
note = {ROC Analysis in Pattern Recognition},
issn = {0167-8655},
doi = {https://doi.org/10.1016/j.patrec.2005.10.010},
url = {https://www.sciencedirect.com/science/article/pii/S016786550500303X},
author = {Tom Fawcett}
}

@inproceedings{binllm,
    title = "Text-like Encoding of Collaborative Information in Large Language Models for Recommendation",
    author = "Zhang, Yang  and
      Bao, Keqin  and
      Yan, Ming  and
      Wang, Wenjie  and
      Feng, Fuli  and
      He, Xiangnan",
    editor = "Ku, Lun-Wei  and
      Martins, Andre  and
      Srikumar, Vivek",
    booktitle = "Proceedings of the 62nd Annual Meeting of the Association for Computational Linguistics (Volume 1: Long Papers)",
    month = aug,
    year = "2024",
    address = "Bangkok, Thailand",
    publisher = "Association for Computational Linguistics",
    url = "https://aclanthology.org/2024.acl-long.497/",
    doi = "10.18653/v1/2024.acl-long.497",
    pages = "9181--9191"
}

@article{collm,
author = {Zhang, Yang and Feng, Fuli and Zhang, Jizhi and Bao, Keqin and Wang, Qifan and He, Xiangnan},
title = {CoLLM: Integrating Collaborative Embeddings Into Large Language Models for Recommendation},
year = {2025},
issue_date = {May 2025},
publisher = {IEEE Educational Activities Department},
address = {USA},
volume = {37},
number = {5},
issn = {1041-4347},
url = {https://doi.org/10.1109/TKDE.2025.3540912},
doi = {10.1109/TKDE.2025.3540912},
journal = {IEEE Trans. on Knowl. and Data Eng.},
month = may,
pages = {2329–2340},
numpages = {12}
}

@article{ndcg,
author = {J\"{a}rvelin, Kalervo and Kek\"{a}l\"{a}inen, Jaana},
title = {Cumulated gain-based evaluation of IR techniques},
year = {2002},
issue_date = {October 2002},
publisher = {Association for Computing Machinery},
address = {New York, NY, USA},
volume = {20},
number = {4},
issn = {1046-8188},
url = {https://doi.org/10.1145/582415.582418},
doi = {10.1145/582415.582418},
journal = {ACM Trans. Inf. Syst.},
month = oct,
pages = {422–446},
numpages = {25}
}

@inproceedings{use-map,
author = {Yue, Yisong and Finley, Thomas and Radlinski, Filip and Joachims, Thorsten},
title = {A support vector method for optimizing average precision},
year = {2007},
isbn = {9781595935977},
publisher = {Association for Computing Machinery},
address = {New York, NY, USA},
url = {https://doi.org/10.1145/1277741.1277790},
doi = {10.1145/1277741.1277790},
booktitle = {Proceedings of the 30th Annual International ACM SIGIR Conference on Research and Development in Information Retrieval},
pages = {271–278},
numpages = {8},
location = {Amsterdam, The Netherlands},
series = {SIGIR '07}
}

@misc{hatllm,
      title={HatLLM: Hierarchical Attention Masking for Enhanced Collaborative Modeling in LLM-based Recommendation}, 
      author={Yu Cui and Feng Liu and Jiawei Chen and Canghong Jin and Xingyu Lou and Changwang Zhang and Jun Wang and Yuegang Sun and Can Wang},
      year={2025},
      eprint={2510.10955},
      archivePrefix={arXiv},
      primaryClass={cs.IR},
      url={https://arxiv.org/abs/2510.10955}, 
}

@InProceedings{sellarec,
author="Wang, Zihan
and Lin, Jinghao
and Yang, Xiaocui
and Liu, Yongkang
and Feng, Shi
and Wang, Daling
and Zhang, Yifei
and Yu, Ge",
editor="Jung, Hyungsoo
and Wang, Tianzheng
and Toyoda, Masashi
and Kwon, Hyuk-Yoon
and Lee, Jae-woong",
title="Enhancing LLM-Based Recommendation with Semantic-Aligned Collaborative Knowledge",
booktitle="Database Systems for Advanced Applications",
year="2026",
publisher="Springer Nature Singapore",
address="Singapore",
pages="405--421",
isbn="978-981-92-0363-5"
}

@ARTICLE{tokenrec,
  author={Qu, Haohao and Fan, Wenqi and Zhao, Zihuai and Li, Qing},
  journal={IEEE Transactions on Knowledge and Data Engineering}, 
  title={TokenRec: Learning to Tokenize ID for LLM-Based Generative Recommendations}, 
  year={2025},
  volume={37},
  number={10},
  pages={6216-6231},
  doi={10.1109/TKDE.2025.3599265}}

@inproceedings{tca4rec,
author = {Lin, Fake and Hu, Binbin and Zheng, Zhi and Zhu, Xi and Liu, Ziqi and Zhang, Zhiqiang and Zhou, Jun and Xu, Tong},
title = {Token-level Collaborative Alignment for LLM-based Generative Recommendation},
year = {2026},
isbn = {9798400723070},
publisher = {Association for Computing Machinery},
address = {New York, NY, USA},
url = {https://doi.org/10.1145/3774904.3792727},
doi = {10.1145/3774904.3792727},
booktitle = {Proceedings of the ACM Web Conference 2026},
pages = {6909–6919},
numpages = {11},
location = {United Arab Emirates},
series = {WWW '26}
}

@misc{ctr-sink,
      title={CTR-Sink: Attention Sink for Language Models in Click-Through Rate Prediction}, 
      author={Zixuan Li and Binzong Geng and Jing Xiong and Yong He and Yuxuan Hu and Jian Chen and Dingwei Chen and Xiyu Chang and Liang Zhang and Linjian Mo and Chengming Li and Chuan Yuan and Zhenan Sun},
      year={2025},
      eprint={2508.03668},
      archivePrefix={arXiv},
      primaryClass={cs.CL},
      url={https://arxiv.org/abs/2508.03668}, 
}

@inproceedings{cora-rec,
author = {Liu, Yuting and Zhang, Jinghao and Dang, Yizhou and Liang, Yuliang and Liu, Qiang and Guo, Guibing and Zhao, Jianzhe and Wang, Xingwei},
title = {CoRA: collaborative information perception by large language model's weights for recommendation},
year = {2025},
isbn = {978-1-57735-897-8},
publisher = {AAAI Press},
url = {https://doi.org/10.1609/aaai.v39i12.33334},
doi = {10.1609/aaai.v39i12.33334},
booktitle = {Proceedings of the Thirty-Ninth AAAI Conference on Artificial Intelligence and Thirty-Seventh Conference on Innovative Applications of Artificial Intelligence and Fifteenth Symposium on Educational Advances in Artificial Intelligence},
articleno = {1361},
numpages = {9},
series = {AAAI'25/IAAI'25/EAAI'25}
}

@inproceedings{gdrt,
author = {Wang, Bohao and Chen, Jiawei and Liu, Feng and Zhang, Changwang and Wang, Jun and Jin, Canghong and Chen, Chun and Wang, Can},
title = {Does LLM Focus on the Right Words? Mitigating Context Bias in LLM-based Recommenders},
year = {2026},
isbn = {9798400723070},
publisher = {Association for Computing Machinery},
address = {New York, NY, USA},
url = {https://doi.org/10.1145/3774904.3792607},
doi = {10.1145/3774904.3792607},
booktitle = {Proceedings of the ACM Web Conference 2026},
pages = {6688–6699},
numpages = {12},
location = {United Arab Emirates},
series = {WWW '26}
}

@ARTICLE{CKF,
  author={Zhao, Chuang and Su, Xing and He, Ming and Zhao, Hongke and Fan, Jianping and Li, Xiaomeng},
  journal={IEEE Transactions on Knowledge and Data Engineering}, 
  title={Collaborative Knowledge Fusion: A Novel Method for Multi-Task Recommender Systems via LLMs}, 
  year={2025},
  volume={37},
  number={9},
  pages={5017-5033},
  doi={10.1109/TKDE.2025.3581706}}

\appendix

\section{Experimental Details}
\label{sec:appendixA}

\subsection{Prompt}
\label{app:prompts}
\begin{figure}[t]
    \centering
    \includegraphics[width=1\columnwidth]{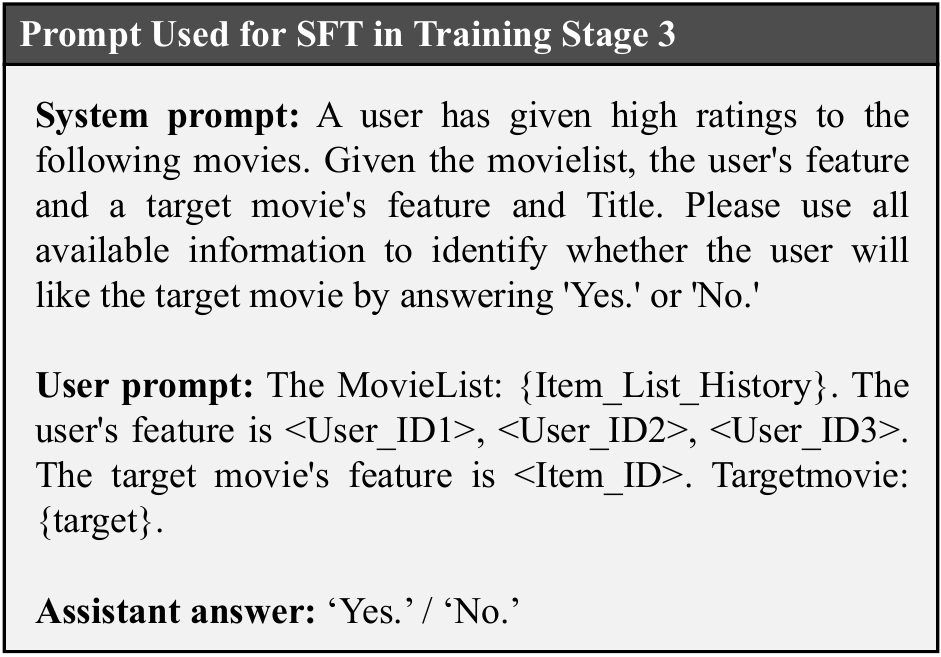}
    \caption{Prompt used for training SAILRec.}
    \label{fig:prompt-sft}
\end{figure}

The prompt used for supervised fine-tuning (SFT) in SAILRec training stage 3 is illustrated in Figure~\ref{fig:prompt-sft}.

\subsection{Codebooks}
\label{app:codebooks}

\begin{figure}[t]
    \centering
    \includegraphics[width=1\columnwidth]{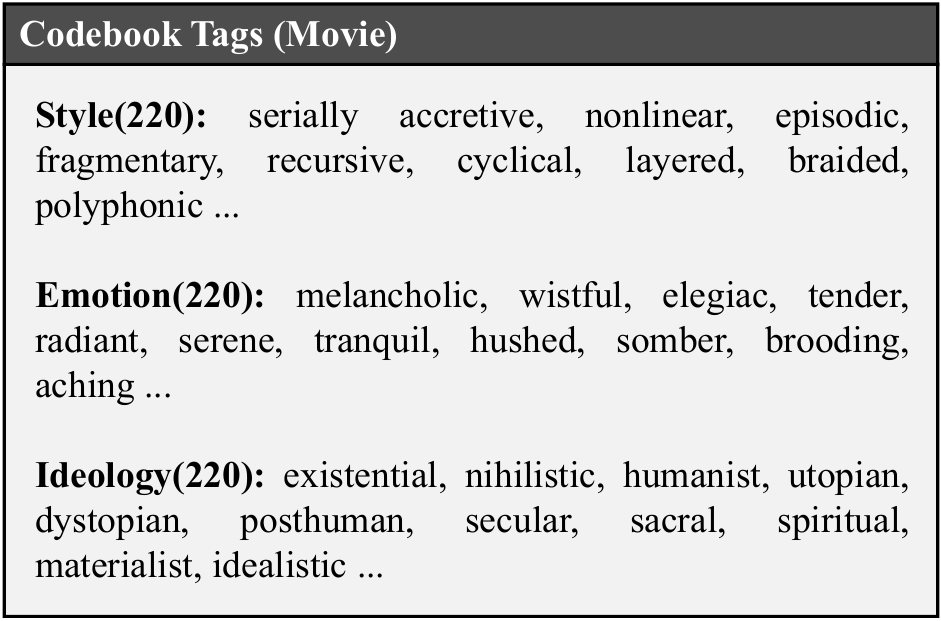}
    \caption{Examples of phrase-level codebook tags for MovieLens-1M. The codebooks cover style, emotion, and ideology, with 220 tags in each semantic dimension.}
    \label{fig:codebook-movie}
\end{figure}

\begin{figure}[t]
    \centering
    \includegraphics[width=1\columnwidth]{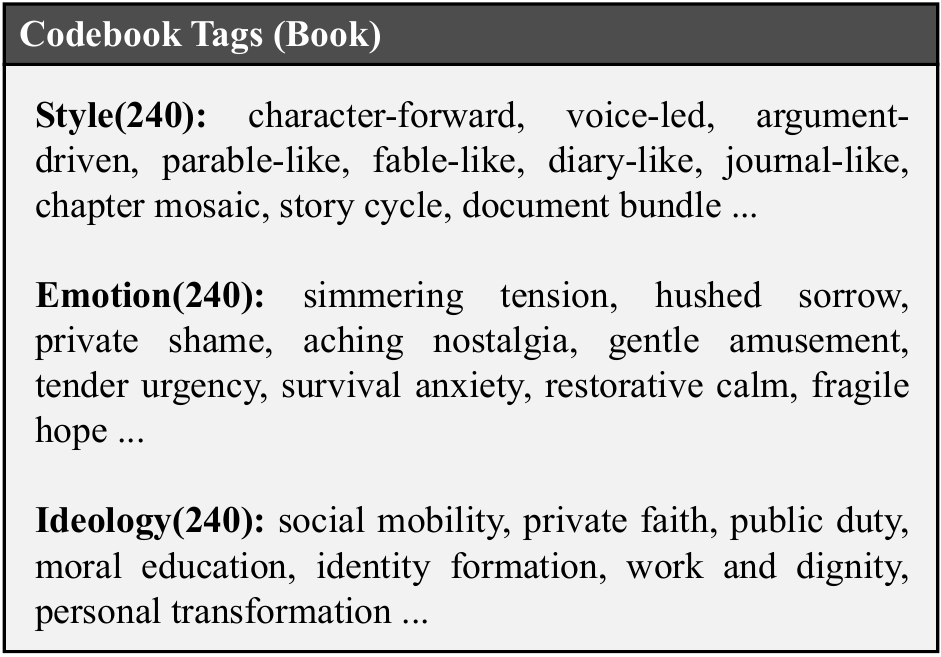}
    \caption{Examples of phrase-level codebook tags for Amazon-Book. The book-domain codebooks are adapted from the movie-domain codebooks and expanded to 240 tags for each semantic dimension.}
    \label{fig:codebook-book}
\end{figure}

To construct semantic alignment targets for user-side collaborative embeddings, we build phrase-level semantic codebooks along three dimensions: style, emotion, and ideology. Each codebook contains tags that describe user preferences from a specific semantic perspective. For MovieLens-1M, each of the three codebooks contains 220 tags. For Amazon-Book, we retain most tags from the movie codebooks and further revise them by removing domain-inappropriate tags and adding book-specific expressions. Each book-domain codebook contains 240 tags. During user-side semantic target construction, historical items are mapped to these tags, and the most representative tags are selected to form user semantic profiles. Partial examples of the codebooks are shown in Figure~\ref{fig:codebook-movie} and Figure~\ref{fig:codebook-book}.

\subsection{Datasets}
\label{app:datasets}

MovieLens-1M~\citep{movielens-1m} is a movie rating dataset released by GroupLens and is one of the most widely used public benchmarks in recommender system research. It contains about one million user-movie ratings collected between 2000 and 2003, involving about 6K users and 4K movies. Each rating ranges from 1 to 5. Following CoLLM and SeLLa-Rec, we convert ratings into binary labels with a threshold of 3, where ratings greater than 3 are treated as positive interactions and the remaining ratings are treated as negative interactions. We keep the interactions from the last 20 months and split them chronologically into training, validation, and test sets with a ratio of 10:5:5. This temporal split evaluates recommendation performance under a realistic setting where future interactions are predicted from earlier histories.

Amazon-Book~\citep{amazon-book} is derived from the book category of Amazon Product Reviews. The original review data contain user ratings, review texts, timestamps, and product metadata, with ratings ranging from 1 to 5. Compared with MovieLens-1M, Amazon-Book contains more users and items, sparser interactions, stronger long-tail distributions, and richer semantic information from book titles and categories. Following prior settings, we use interactions from 2017 and keep users with more than 20 interactions to ensure data quality. Review scores are converted into binary labels with a threshold of 4. The data are then sorted by timestamp and split chronologically into training, validation, and test sets with a ratio of 11:0.5:0.5.

After preprocessing, both datasets are organized into click-through rate prediction samples, where each instance consists of a user historical interaction sequence, a candidate item, and a binary label. The processed statistics are shown in Table~\ref{tab:dataset_statistics}.

\begin{table}[t]
\centering
\scriptsize
\setlength{\tabcolsep}{7pt}
\begin{tabular}{@{}lrrrrr@{}}
\toprule
Dataset & \#Train & \#Valid & \#Test & \#User & \#Item \\
\midrule
MovieLens-1M & 33,891 & 10,401 & 7,331 & 839 & 3,256 \\
Amazon-Book & 727,468 & 25,747 & 25,747 & 22,967 & 34,154 \\
\bottomrule
\end{tabular}
\caption{Statistics of datasets.}
\label{tab:dataset_statistics}
\end{table}

\subsection{Baselines}
\label{app:baselines}

To comprehensively evaluate SAILRec, we compare it with three groups of representative baselines, including conventional collaborative filtering models, LLM-only recommenders, and collaborative-enhanced LLM recommenders.

Collaborative filtering models:

$\bullet$ MF~\citep{mf} is a classic matrix factorization method. It learns low-dimensional latent vectors for users and items, and estimates their matching score through the inner product between the two vectors.

$\bullet$ LightGCN~\citep{lightgcn} is a graph-based collaborative filtering model built on the user-item interaction graph. It removes feature transformation and nonlinear activation from conventional GCNs, and keeps only neighborhood aggregation to capture high-order collaborative signals.

$\bullet$ SASRec~\citep{sasrec} is a representative sequential recommendation model. It uses self-attention to model dynamic user interests from historical interaction sequences and predicts user preferences for candidate items.

LLM-only recommenders:

$\bullet$ TALLRec~\citep{tallrec} constructs recommendation samples as instruction data and performs lightweight tuning to align LLMs with recommendation tasks. It enables LLMs to directly generate preference judgments such as \texttt{Yes} or \texttt{No}.

$\bullet$ GDRT~\citep{gdrt} studies context bias in LLM-based recommendation, where supervised fine-tuning may over-rely on auxiliary tokens and underuse user interaction tokens. It adopts Group DRO to improve robustness across token relevance groups.

$\bullet$ HatLLM~\citep{hatllm} analyzes the limitation of LLM attention in modeling cross-item collaborative relations. It introduces hierarchical attention masking to strengthen intra-item semantic understanding in shallow layers and cross-item relation modeling in deep layers.

Collaborative-enhanced LLM recommenders:

$\bullet$ CoLLM~\citep{collm} treats collaborative information as an additional modality. It extracts user and item collaborative embeddings from an external collaborative model and maps them into the LLM input embedding space as soft prompts.

$\bullet$ BinLLM~\citep{binllm} represents collaborative information in a text-like form. It converts collaborative embeddings from external models into binary sequences, with an optional dot-decimal compression, so that LLMs can process collaborative knowledge as textual features.

$\bullet$ CoRA~\citep{cora-rec} aligns collaborative information with the parameter space of LLMs instead of the input space. It converts collaborative queries into low-rank incremental weights, enabling LLMs to perceive collaborative signals without inserting extra collaborative tokens into prompts.

$\bullet$ CKF~\citep{CKF} targets multi-task recommendation with collaborative knowledge fusion. It maps collaborative embeddings into LLM prompts through personalized mapping functions and uses Multi-LoRA to separate task-shared and task-specific information.

$\bullet$ TokenRec~\citep{tokenrec} focuses on user and item ID tokenization for LLM-based recommendation. It quantizes collaborative representations into discrete tokens through a masked vector-quantized tokenizer and uses generative retrieval for efficient recommendation.

$\bullet$ SeLLa-Rec~\citep{sellarec} addresses the semantic gap between collaborative knowledge and LLM semantic knowledge. It uses semantic distillation, contrastive alignment, and a hybrid projection layer to inject semantic-aligned collaborative knowledge into LLMs.

$\bullet$ TCA4Rec~\citep{tca4rec} incorporates collaborative signals at the optimization level. It converts item-level collaborative filtering logits into token-level distributions and combines them with one-hot labels to form soft supervision for next-token prediction.

\subsection{Evaluation Metrics}
\label{app:metrics}

Under the click-through rate prediction setting, we use the predicted positive-class probability as the candidate item score. Specifically, for each user-item pair $(u,i)$, the LLM produces logits for \texttt{Yes} and \texttt{No} at the answer position. Let $o_{\text{Yes}}$ and $o_{\text{No}}$ denote the corresponding logits. The prediction score is computed by a binary softmax:
\begin{equation}
s_{u,i}
=
\frac{\exp(o_{\text{Yes}})}
{\exp(o_{\text{Yes}})+\exp(o_{\text{No}})}.
\end{equation}
This score is used for all evaluation metrics. In implementation, the logits of \texttt{Yes} and \texttt{No} are first extracted from the vocabulary distribution, normalized with a binary softmax, and the probability of \texttt{Yes} is used as the positive-class score.

AUC evaluates the overall discrimination ability between positive and negative instances over all test samples. Let $\mathcal{D}$ denote the test set, and let $\mathcal{D}^{+}$ and $\mathcal{D}^{-}$ denote the positive and negative subsets, respectively. AUC is computed as
\begin{equation}
\mathrm{AUC}
=
\frac{1}{|\mathcal{D}^{+}||\mathcal{D}^{-}|}
\sum_{p\in\mathcal{D}^{+}}
\sum_{n\in\mathcal{D}^{-}}
\psi(s_p,s_n),
\label{eq:auc}
\end{equation}
where
\[
\psi(a,b)=
\begin{cases}
1, & a>b,\\
\frac{1}{2}, & a=b,\\
0, & a<b.
\end{cases}
\]
Here, $\psi(a,b)$ assigns a score of $1$ when a positive instance receives a higher prediction score than a negative instance, $\frac{1}{2}$ when the two scores are tied, and $0$ otherwise. In our implementation, global AUC is computed with \texttt{roc\_auc\_score} over all test labels and prediction scores.

UAUC evaluates ranking performance at the user level. For each user $u$, we collect the candidate set $\mathcal{D}_u=\{(i,y_{u,i},s_{u,i})\}$ from the test set and compute AUC within this user. Users with only one candidate sample or with only one label class are excluded, because AUC is undefined in these cases. Let $\mathcal{U}_{\mathrm{auc}}$ denote the remaining valid users. UAUC is then computed as the unweighted mean of per-user AUC:
\begin{equation}
\mathrm{UAUC}
=
\frac{1}{|\mathcal{U}_{\mathrm{auc}}|}
\sum_{u\in \mathcal{U}_{\mathrm{auc}}}
\mathrm{AUC}_u .
\end{equation}
Different from impression-weighted user AUC, each valid user contributes equally in this metric. Therefore, UAUC reduces the influence of user activity imbalance and better reflects personalized discrimination ability across users.

NDCG evaluates whether positive items are ranked near the top of each user's candidate list. For each valid user $u$, candidate items are sorted in descending order according to $s_{u,i}$. Let $y_{u,r}\in\{0,1\}$ denote the binary relevance label at rank $r$ after sorting, and let $n_u$ be the number of candidates for user $u$. The discounted cumulative gain is
\begin{equation}
\mathrm{DCG}_u
=
\sum_{r=1}^{n_u}
\frac{y_{u,r}}{\log_2(r+1)}.
\end{equation}
Let $P_u=\sum_{r=1}^{n_u} y_{u,r}$ denote the number of positive items for user $u$. The ideal discounted cumulative gain is
\begin{equation}
\mathrm{IDCG}_u
=
\sum_{r=1}^{P_u}
\frac{1}{\log_2(r+1)}.
\end{equation}
The user-level NDCG is then
\begin{equation}
\mathrm{NDCG}_u
=
\frac{\mathrm{DCG}_u}{\mathrm{IDCG}_u}.
\end{equation}
In implementation, users without both positive and negative samples are skipped to avoid invalid ranking evaluation. The final NDCG is computed as the unweighted mean over all valid users:
\begin{equation}
\mathrm{NDCG}
=
\frac{1}{|\mathcal{U}_{\mathrm{rank}}|}
\sum_{u\in \mathcal{U}_{\mathrm{rank}}}
\mathrm{NDCG}_u ,
\end{equation}
where $\mathcal{U}_{\mathrm{rank}}$ denotes the set of users used for ranking evaluation.

MAP measures the average precision of positive items in the ranked candidate list. For each valid user $u$, after sorting candidate items by prediction score, precision at rank $r$ is defined as
\begin{equation}
\mathrm{Prec}_u(r)
=
\frac{1}{r}
\sum_{t=1}^{r} y_{u,t}.
\end{equation}
The average precision for user $u$ is
\begin{equation}
\mathrm{AP}_u
=
\frac{1}{P_u}
\sum_{r=1}^{n_u}
y_{u,r}\mathrm{Prec}_u(r),
\end{equation}
where $P_u$ is the number of positive items for this user. The final MAP is the unweighted mean of user-level average precision:
\begin{equation}
\mathrm{MAP}
=
\frac{1}{|\mathcal{U}_{\mathrm{rank}}|}
\sum_{u\in \mathcal{U}_{\mathrm{rank}}}
\mathrm{AP}_u .
\end{equation}
Thus, AUC measures global discrimination over all samples, while UAUC, NDCG, and MAP evaluate user-level personalized ranking quality by first computing metrics within each user and then averaging over valid users.

\subsection{Training and Hyperparameter Details}
\label{app:other_settings}

SAILRec uses Qwen2-7B as the backbone LLM. For fair comparison, all LLM-based baseline methods are also evaluated with Qwen2-7B as the backbone. In the second training stage, SAILRec performs dual-side semantic alignment with contrastive learning. The learning rate is set to $1\times10^{-3}$, the batch size is set to 128, and the temperature in the InfoNCE loss is set to 0.07.

In the third training stage, SAILRec performs supervised fine-tuning (SFT) for click-through rate prediction. The learning rate is selected from the range between $8\times10^{-5}$ and $2\times10^{-4}$. We set the weight decay to 0.01, $\beta_2$ in Adam to 0.95, the warmup ratio to 0, and the maximum sequence length to 768. For MovieLens-1M, the batch size is set to 256 and the model is trained for 2 epochs. For Amazon-Book, the batch size is set to 2000 and the model is trained for 1 epoch. All experiments are conducted on NVIDIA A6000 GPUs. The total training time is about 7 hours on MovieLens-1M and about 92 hours on Amazon-Book. For all experimental results, we run each model multiple times and report the test performance of the checkpoint with the best overall validation performance.

\section{Additional Experimental Results and Analysis}
\label{sec:appendixB}

\subsection{Additional Analysis of Collab. tokens}
\label{app:add-collab-analysis}

\begin{figure}[t]
    \centering
    \includegraphics[width=1\columnwidth]{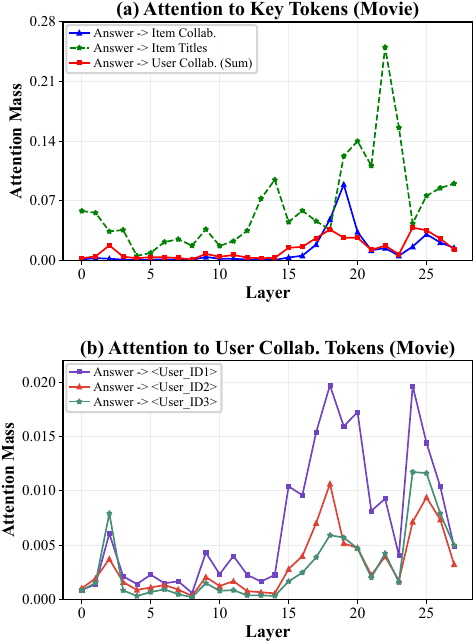}
    \caption{Layer-wise answer attention to semantic and collaborative tokens in SAILRec on MovieLens-1M, with user-side attention further decomposed into three embeddings.}
    \label{fig:attention_key_tokens}
\end{figure}

\begin{figure}[t]
    \centering
    \includegraphics[width=1\columnwidth]{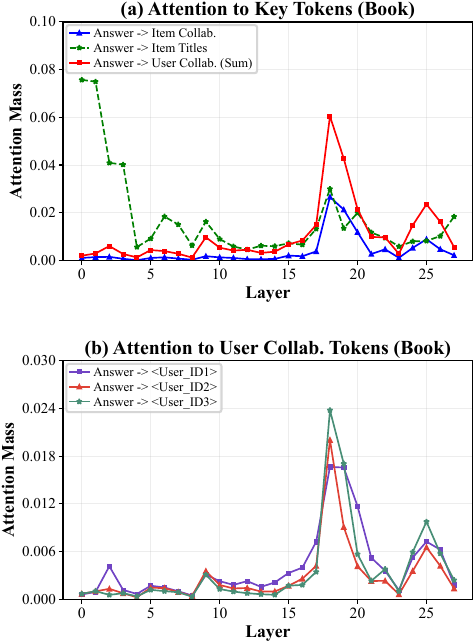}
    \caption{Layer-wise answer attention to semantic and collaborative tokens in SAILRec on Amazon-Book. The upper panel compares attention to item titles, item-side Collab. tokens, and summed user-side Collab. tokens, while the lower panel decomposes the attention to the three user-side Collab. tokens.}
    \label{fig:att-sailrec-book}
\end{figure}

\begin{figure*}[!t]
    \centering

    \includegraphics[width=0.48\textwidth]{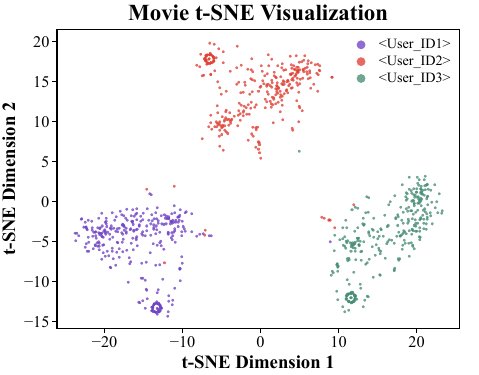}
    \hfill
    \includegraphics[width=0.48\textwidth]{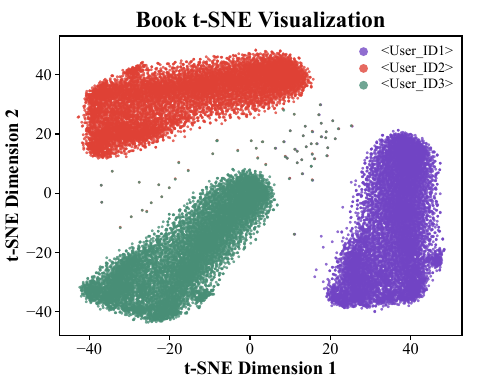}
    \caption{\textit{t}-SNE visualization of the three user-side Collab. tokens on MovieLens-1M and Amazon-Book. Different colors denote the three semantic directions learned by the user-side C-QFormer.}
    \label{fig:tsne}

    \vspace{1.0em}

    \includegraphics[width=0.48\textwidth]{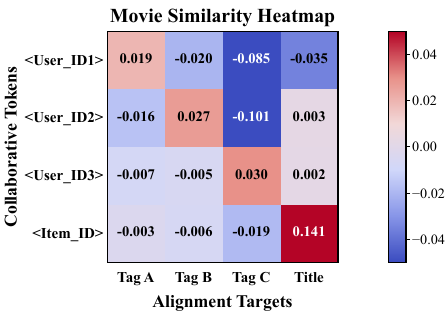}
    \hfill
    \includegraphics[width=0.48\textwidth]{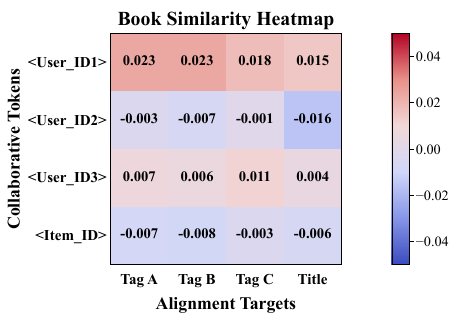}
    \caption{Similarity heatmaps between learned Collab. tokens and semantic alignment targets on MovieLens-1M and Amazon-Book. Rows correspond to user-side and item-side Collab. tokens, and columns correspond to codebook tags and item-title semantic targets.}
    \label{fig:heat-map}
\end{figure*}

We further analyze how SAILRec uses collaborative embeddings inside the LLM. Figure~\ref{fig:attention_key_tokens} shows the layer-wise attention patterns on MovieLens-1M. The attention to Collab. tokens remains low in shallow layers, which is consistent with our design to avoid premature interference with the understanding of historical item tokens. It then increases in middle and deep layers, indicating that collaborative evidence is gradually introduced near the decision stage. Meanwhile, the attention to Collab. tokens does not exceed that to historical item-title tokens, suggesting that SAILRec enhances collaborative evidence without overusing it. For the three user-side Collab. tokens, their attention curves follow similar depth-wise trends, which also matches the hierarchical steering design.

Figure~\ref{fig:att-sailrec-book} reports the layer-wise attention patterns on Amazon-Book. Similar to MovieLens-1M, the attention to Collab. tokens stays low in shallow layers and becomes more active in middle-to-deep layers. This confirms that SAILRec does not force collaborative evidence to dominate the whole inference process, but introduces it mainly when preference matching becomes more relevant. A difference is that Amazon-Book shows sharper attention peaks around the decision-related layers, especially for summed user-side Collab. tokens. This may be related to the stronger sparsity and semantic diversity of book interactions, where user-side preference abstraction becomes more important than relying only on item titles. In contrast, MovieLens-1M exhibits stronger and more persistent attention to historical item-title tokens, which is reasonable because movie titles and genres often provide more compact semantic cues. In both datasets, the three user-side Collab. tokens follow similar depth-wise trends, but their magnitudes differ, suggesting that they participate in the same inference stage while capturing complementary preference semantics.

Figure~\ref{fig:tsne} further visualizes the learned user-side Collab. tokens with \textit{t}-SNE. The visualization suggests that the three user-side Collab. tokens occupy distinguishable regions in the projected space on both datasets. This pattern indicates that the user-side C-QFormer does not simply generate identical or fully redundant representations for the three placeholders. Instead, the mapped embeddings tend to preserve different representation patterns that correspond to the three semantic directions used in user-side alignment. The separation is more compact on Amazon-Book, which may be related to its larger data scale and richer phrase-level preference space.

Figure~\ref{fig:heat-map} shows the similarity between learned Collab. tokens and their semantic alignment targets. On MovieLens-1M, the item-side Collab. token shows a clearer association with item-title semantics, suggesting that item-side collaborative representations can be more directly grounded in LLM semantic anchors when item titles and genres provide stable semantic cues. The user-side tokens show weaker but differentiated associations with their corresponding codebook targets, which is reasonable because user preference profiles are aggregated from multiple historical interactions rather than derived from a single textual description. On Amazon-Book, the similarities are generally weaker and less concentrated, while user-side tokens exhibit broader associations with multiple semantic targets. This may reflect the larger item space, sparser interactions, and more diverse book semantics, where a single title representation may not fully capture the collaborative meaning of an item. Overall, these observations provide qualitative evidence that dual-side semantic alignment can induce different semantic roles for user-side and item-side Collab. tokens.

\begin{table*}[t]
\centering
\footnotesize
\setlength{\tabcolsep}{5.0pt}
\renewcommand{\arraystretch}{1.12}
\begin{tabular*}{\textwidth}{@{\extracolsep{\fill}}lcccccccc@{}}
\toprule
\multirow{2}{*}{\textbf{Method}}
& \multicolumn{4}{c}{\textbf{MovieLens-1M}}
& \multicolumn{4}{c}{\textbf{Amazon-Book}} \\
\cmidrule(lr){2-5} \cmidrule(lr){6-9}
& \textbf{AUC} & \textbf{UAUC} & \textbf{NDCG} & \textbf{MAP}
& \textbf{AUC} & \textbf{UAUC} & \textbf{NDCG} & \textbf{MAP} \\
\midrule
SAIL-mask-U
& 0.7248 & 0.7008 & 0.8797 & 0.7670
& 0.5553 & 0.5694 & 0.8281 & 0.7477 \\
SAIL-mask-I
& 0.7074 & 0.6613 & 0.8637 & 0.7415
& 0.8168 & 0.7136 & 0.8866 & 0.8306 \\
SAIL-mask-UI
& 0.6941 & 0.6758 & 0.8687 & 0.7494
& 0.5361 & 0.5452 & 0.8178 & 0.7337 \\
SAIL-mask-Shallow
& 0.6750 & 0.6364 & 0.8483 & 0.7198
& 0.6944 & 0.6532 & 0.8594 & 0.7921 \\
SAIL-mask-Middle
& 0.7166 & 0.6700 & 0.8601 & 0.7397
& 0.7178 & 0.6600 & 0.8662 & 0.8012 \\
SAIL-mask-Deep
& 0.7541 & 0.6840 & 0.8689 & 0.7569
& 0.8247 & 0.7374 & 0.8954 & 0.8431 \\
\midrule
\textbf{Full-SAILRec}
& \textbf{0.7571} & \textbf{0.7410} & \textbf{0.8985} & \textbf{0.7999}
& \textbf{0.8393} & \textbf{0.7506} & \textbf{0.9017} & \textbf{0.8524} \\
\bottomrule
\end{tabular*}
\caption{Inference-time attention bias masking results on MovieLens-1M and Amazon-Book. SAIL-mask-U, SAIL-mask-I, and SAIL-mask-UI mask the attention paths to user-side, item-side, and both-side Collab. tokens, respectively. SAIL-mask-Shallow, SAIL-mask-Middle, and SAIL-mask-Deep mask the attention paths to all Collab. tokens in the corresponding layer groups.}
\label{tab:bias_masking}
\end{table*}
\subsection{Inference-time Attention Bias Masking}
\label{app:attention_bias_masking}

\begin{table*}[t]
\centering
\footnotesize
\setlength{\tabcolsep}{5.0pt}
\renewcommand{\arraystretch}{1.12}
\begin{tabular*}{\textwidth}{@{\extracolsep{\fill}}lcccccccc@{}}
\toprule
\multirow{2}{*}{\textbf{Method}}
& \multicolumn{4}{c}{\textbf{MovieLens-1M}}
& \multicolumn{4}{c}{\textbf{Amazon-Book}} \\
\cmidrule(lr){2-5} \cmidrule(lr){6-9}
& \textbf{AUC} & \textbf{UAUC} & \textbf{NDCG} & \textbf{MAP}
& \textbf{AUC} & \textbf{UAUC} & \textbf{NDCG} & \textbf{MAP} \\
\midrule
CoLLM
& 0.6916 & 0.6666 & 0.8665 & 0.7502
& 0.7520 & 0.6874 & 0.8750 & 0.8131 \\
SeLLa-Rec
& 0.7101 & 0.6938 & 0.8780 & 0.7657
& 0.8036 & 0.7152 & 0.8863 & 0.8310 \\
\midrule
SAIL-w/o-S
& 0.7176 & 0.6870 & 0.8579 & 0.7397
& 0.8165 & 0.7247 & 0.8920 & 0.8382 \\
SAIL-w/o-UI
& 0.7057 & 0.6973 & 0.8719 & 0.7569
& 0.8042 & 0.7020 & 0.8801 & 0.8220 \\
\midrule
Full-SAILRec
& \textbf{0.7398} & \textbf{0.7114} & \textbf{0.8861} & \textbf{0.7784}
& \textbf{0.8359} & \textbf{0.7407} & \textbf{0.8986} & \textbf{0.8485} \\
\bottomrule
\end{tabular*}
\caption{Performance comparison on MovieLens-1M and Amazon-Book using Qwen3-8B as the backbone.}
\label{tab:qwen3_results}
\end{table*}

To further examine the relation between attention bias and recommendation performance, we conduct an inference-time attention bias masking experiment. Unlike retraining-based ablations, this experiment keeps all Full-SAILRec parameters fixed and only intervenes in specific attention paths during the forward pass. Specifically, we set the attention bias of the corresponding Collab. token positions to $-\infty$, preventing the model from attending to these tokens. Thus, the resulting performance drop provides interventional evidence for the contribution of the masked paths.

Table~\ref{tab:bias_masking} includes side-wise and layer-wise masking variants. SAIL-mask-U, SAIL-mask-I, and SAIL-mask-UI mask the user-side, item-side, and both-side Collab. tokens, respectively. SAIL-mask-Shallow, SAIL-mask-Middle, and SAIL-mask-Deep mask the attention paths to all Collab. tokens in the corresponding layer groups.

All masking variants underperform Full-SAILRec, showing that Collab. tokens contribute to prediction rather than merely receiving attention in visualization. Side-wise masking reveals complementary roles of user-side and item-side collaborative signals. On Amazon-Book, masking user-side Collab. tokens causes a large AUC drop from $0.8393$ to $0.5553$, suggesting that user-side preference abstraction is crucial in the sparse and long-tailed book domain. On MovieLens-1M, masking item-side Collab. tokens also causes clear degradation, indicating the importance of item-side collaborative evidence for candidate discrimination. SAIL-mask-UI usually suffers the largest drop, further showing that both sides provide complementary information.

Layer-wise masking shows that Collab. tokens contribute unequally across Transformer depths. Masking shallow layers causes a large drop, which does not contradict shallow-layer suppression because SAILRec weakens, rather than removes, early collaborative access. Masking middle layers also degrades performance, suggesting that middle layers mediate interactions between textual semantics and collaborative information. When deep layers are masked, global AUC sometimes drops less, but UAUC, NDCG, and MAP still decrease notably, indicating that deep-layer collaborative attention is more related to personalized ranking and final preference matching. Overall, the results provide stronger evidence than attention visualization alone and support the design of dual-side semantic alignment and hierarchical attention steering.

\subsection{Performance on Qwen3-8B}
\label{app:qwen3_results}

To examine backbone generalization, we further conduct experiments with Qwen3-8B~\citep{qwen3} in the no-think mode. For controlled comparison, we reuse the Qwen2-7B training hyperparameters, including the learning rate range, LoRA configuration, batch setting, and attention steering schedule. As shown in Table~\ref{tab:qwen3_results}, performance on Qwen3-8B is generally lower than on Qwen2-7B, although the decline is not uniform. This may result from insufficient backbone-specific tuning, as Qwen3-8B may have different representation spaces, logit distributions, and layer-wise attention dynamics. Its stronger semantic prior may also reduce the use of external heterogeneous collaborative information, especially without semantic alignment. Thus, the lower scores suggest the need for backbone-specific tuning and semantic alignment, rather than failure of the method.

Despite the drop, the relative results still support SAILRec. Here, SAIL-w/o-S removes hierarchical attention steering, and SAIL-w/o-UI removes user-side and item-side semantic alignment. SeLLa-Rec improves over CoLLM on all metrics, showing that semantic alignment remains useful for Qwen3-8B. Full-SAILRec achieves the best results across both datasets and all metrics. The comparison with SAIL-w/o-UI indicates that dual-side semantic alignment helps beyond direct collaborative embedding injection, while the comparison with SAIL-w/o-S shows that attention steering improves layer-wise utilization. On Amazon-Book, Full-SAILRec brings clear UAUC, NDCG, and MAP gains over both ablation variants, suggesting that the two designs are especially helpful in the sparser book domain. Overall, SAILRec remains effective under Qwen3-8B without backbone-specific hyperparameter tuning.

\end{document}